\documentclass[a4paper,12pt]{article}
\usepackage{amsmath, amssymb,graphicx}
\usepackage{braket}
\pdfoutput=1
\usepackage{jheppub}

\newcommand{\beq}{\begin{equation}}
\newcommand{\eeq}{\end{equation}}
\newcommand{\bea}{\vspace{0.25cm}\begin{eqnarray}}
\newcommand{\eea}{\end{eqnarray}}

\newcommand{\ro}{\mbox{{\boldmath
$\rho$}}}

\newcommand{\pb}{{{\bf p}}}

\newcommand{\bb}{{{\bf b}}}

\def\lsim{\mathrel{\rlap{\lower4pt\hbox{\hskip1pt$\sim$}}
    \raise1pt\hbox{$<$}}}         
\def\gsim{\mathrel{\rlap{\lower4pt\hbox{\hskip1pt$\sim$}}
    \raise1pt\hbox{$>$}}}         

\author{B.G.~Zakharov$^{1,2}$}

\affiliation{$^{1}$L.D.~Landau Institute for Theoretical Physics,
        GSP-1, 117940, Kosygina Str. 2, 117334 Moscow, Russia}
\affiliation{$^{2}$Steklov Mathematical Institute, Russian Academy of Sciences,
Gubkin str. 8, 119991 Moscow, Russia}

\emailAdd{bgz@itp.ac.ru}

\begin{document}

\title{
  Jet quenching from heavy to light ion collisions  
}

\abstract{  
  We perform an analysis of jet quenching in heavy and light ion collisions
  for scenarios without and with
  quark-gluon plasma formation in $pp$ collisions.
  We find that the results for these scenarios are very similar,
  and both of them are in reasonable
  agreement with data for heavy ion collisions.
  However, their results become differ significantly for light nuclei.
  Using the parameters fitted to heavy ion data on the nuclear
  modification factor $R_{AA}$, we make predictions
  for 0.2 and 7 TeV O+O collisions that can be verified by
  future experiments at RHIC and the LHC.
}


\maketitle

\section{Introduction}
It is believed that heavy ion collisions at the RHIC and LHC energies
produce a deconfined quark-gluon plasma (QGP). Strong suppression of
the high-$p_T$ particles observed in heavy ion collisions
(usually called jet quenching) at RHIC and the LHC
is one of the major signals of the QGP formation.
This phenomenon is a consequence of parton energy loss in the QGP.
For fast partons the energy loss is dominated by the radiative mechanism
through induced gluon emission \cite{GW,BDMPS1,LCPI1,W1,GLV1,AMY1}.
The suppression of particle spectra
in $AA$ collisions is quantified by the
nuclear modification factor $R_{AA}$. For a given
centrality bin $\Delta c$ (which is experimentally determined
via charged hadron multiplicities) $R_{AA}$ is
defined as the ratio of particle yield in $AA$ collisions
to that in $pp$ collisions scaled with the number of binary
collisions
\beq
R_{AA}=\frac{d^2N_{AA}/dp_T^2dy}{N_{ev}
  \langle T_{AA}\rangle_{\Delta c} d^2\sigma_{pp}/dp_T^2dy},
\label{eq:10}
\eeq
where $\langle T_{AA}\rangle_{\Delta c}$ is the nuclear overlap function
for centrality class $\Delta c$. The relation between centrality $c$
and the impact parameter $b$ of $AA$ collision is usually defined
within the Glauber model.
Theoretical calculations of the nuclear suppression
are commonly performed assuming that the medium jet modification occurs only
in $AA$ collisions, and is absent in $pp$ collisions, i.e.,
the $pp$ yield in the denominator on the righthand side of (\ref{eq:10})
can be calculated in the standard pQCD approach.
In this scenario, with the Glauber initial conditions for the QGP fireball
at the proper time $\tau_0\sim 0.5$ fm, one can obtain
within pQCD jet quenching models rather good description of the
available data on the
nuclear modification factor from RHIC for Au+Au collisions and from LHC for Pb+Pb and Xe+Xe
collisions (see, e.g., Refs.
\cite{CUJET3,Djord1,Armesto_LHC,W_mQGP1,RAA20}).
However, it is possible that the assumption that QGP is not formed
in $pp$ collisions is not valid. The observation of the ridge effect
in high-multiplicity  $pp$ collisions at the LHC \cite{CMS_ridge}
(and even in lower-multiplicity $pp$ collisions \cite{ATLAS_mbias}), suggests
that a small size QGP fireball can be created in $pp$ collisions as well,
because the hydrodynamic transverse flow seems to
be the most natural explanation for the ridge effect.
But an alternative interpretation of the ridge phenomenon
via the initial state effects \cite{ridge_CGC},
without the collective flow in the final state, is not excluded.
The formation of the QGP droplets in collisions of small systems is now one
of the challenging issues for high energy QCD.
Future experiments on light ion collisions at RHIC \cite{CERN_STAR}
and at the LHC \cite{CERN_yel,CERN_LHC,CERN_OOsum}
will allow to understand better the role of the collective
effects in small systems.

In the case of a mini-QGP (mQGP) formation in $pp$ collisions,
even if it occurs only in a part of jet events,
the parton shower and the inclusive $pp$ cross section will be affected by
the parton energy loss in the mQGP.
It is important that the jet events in $pp$ collisions
have favorable conditions for QGP formation, as compared to
the minimum bias events, since 
the charged multiplicity density
of the soft (underlying-event (UE)) hadrons in the jet events,
$dN_{ch}^{ue}/d\eta$, turns out to be bigger than the ordinary minimum bias
multiplicity density by
a factor (we denote it as $K_{ue}$) of $\sim 2-2.5$ \cite{Field}.
The $K_{ue}$ grows with momentum of the 
leading charged jet hadron at $p_{T}\lsim 3-5$ GeV, and then flattens out
\cite{CDF,PHENIX_dA,ATLAS_UE_Nch,CMS_UE_Nch,ALICE_UE_Nch}.
The enhancement of the UE multiplicity may be due to the fact
that jet production should be most probable for central $pp$ collisions,
and due to the initial state radiation that accompanies hard jet production.
For the LHC energies $dN_{ch}^{ue}/d\eta$ at $\eta\sim 0$
turns out to be larger than the critical
pseudorapidity density $dN_{ch}/d\eta \sim 6$   for
beginning of the regime with formation of the hot QCD matter
in $pp$ collisions found in \cite{Camp1} from variation of the average
transverse hadron momentum $\langle p_T\rangle$ with $dN_{ch}/d\eta$,
and than a more conservative value
$dN_{ch}/d\eta \sim 9$ obtained
in \cite{Csanad}\footnote{
  We make a reasonable assumption that
for $dN_{ch}^{ue}/d\eta$ equal to $dN_{ch}/d\eta$
  dynamics of the UE soft hadron
production in $pp$ jet events is similar to that for $pp$ collisions
without jet detecting.}.
The onset of the regime with QGP formation
at $dN_{ch}/d\eta\gsim 5-10$ is also supported by the observation
of a steep growth  of the strange particle
production in $pp$ collisions 
in the range $dN_{ch}/d\eta\sim 2-8$ \cite{ALICE_strange}.
For $pp$ collisions at the RHIC energy $\sqrt{s}=0.2$ TeV
$dN_{ch}^{ue}/d\eta\sim 6$. Thus, one can expect for the UEs in jet
production in $pp$ collisions at RHIC energies the thermalization may be
incomplete.
Of course, the transition to the regime of a thermalized
QCD matter may be smooth. It is possible that at the beginning
the final state interactions in the parton system
are weak and lead only to a small modification of the free-stream evolution
\cite{Kurkela}. Also, it is clear that, due to event-by-event fluctuations of
the entropy/energy production, the QGP can be formed in some part of
jet events   even for
a small average UE multiplicity density at $\sqrt{s}=0.2$ TeV.
Therefore, some medium jet modification in $pp$ collisions may already
occur at the RHIC energies.

In the presence of the mQGP, the real inclusive $pp$ cross section
is determined by the product of the pQCD cross section and the medium
modification factor $R_{pp}$ (which accounts for the jet final state
interaction in the mQGP)
\beq
d\sigma(pp\to hX)/d\pb_{T}dy=
R_{pp} d\sigma_{pert}(pp\to hX)/d\pb_{T}dy\,.
\label{eq:20}
\eeq
The first pQCD calculations of $R_{pp}$
have been performed in \cite{Z_pp_PRL,Z_pp13}  within the light-cone path
integral (LCPI) approach \cite{LCPI1} to induced gluon emission,
using the measured UE charged multiplicity density
for calculation of the initial mQGP temperature.
The results of \cite{Z_pp13} show that for $pp$ collisions at the LHC energies
medium suppression of hadron spectra can be $\sim 20-30$\%
at $p_T\sim 10-20$ GeV. Since $R_{pp}$ is rather close to unity and
has a smooth $p_T$-dependence it is difficult to differentiate
the scenarios with and without medium suppression using only data
on the high-$p_T$ hadron spectra in $pp$ collisions.
In \cite{Z_pp_PRL} it was suggested to use as a signal of jet quenching in $pp$
collisions variation of the photon/hadron-tagged jet
fragmentation functions (FFs) with the UE multiplicity.
The recent preliminary data from ALICE \cite{ALICE_Ipp} on
the jet FF modification factor $I_{pp}$ at $\sqrt{s}=5.02$ TeV
for the hadron-tagged jets (with the trigger hadron momentum $8<p_T<15$ GeV,
and the associated away side hadron
momentum in the range  $4<p_T<6$ GeV) show
a monotonic decrease of $I_{pp}$ with the UE multiplicity
by about 20\% for the UE multiplicity density range $\sim 5-20$.
This agrees qualitatively with predictions  of \cite{Z_pp_PRL}.
The observation of decrease of $I_{pp}$ with the UE multiplicity, if confirmed,
will be a strong argument for the scenario with mQGP formation
in $pp$ jet events. 

Evidently, for the scenario with mQGP formation in $pp$ collisions 
the theoretical $R_{AA}$, which corresponds to the experimentally measured
$R_{AA}$, given by (\ref{eq:10}), reads
\beq
R_{AA}=R_{AA}^{st}/R_{pp}\,,
\label{eq:30}
\eeq
where $R_{AA}^{st}$ is the standard  nuclear modification factor 
calculated with the pQCD $pp$ cross section.
For a weak jet quenching in the mQGP, i.e., for $R_{pp}$ close to unity,
observation of the effect of the $1/R_{pp}$ factor in (\ref{eq:30})
from comparison with
data on $R_{AA}$ for heavy nuclei, when $R_{AA}\sim 0.15-0.5$,
is a hopeless task, because uncertainties
of the present theoretical jet quenching models are rather large and the
presence of $1/R_{pp}$ in (\ref{eq:30}) may be mimicked 
by changes of the parameter values.
However, the effect from $1/R_{pp}$ factor can
become more pronounced in the case of collisions
of light nuclei, when $R_{AA}^{st}$ also becomes close to unity.
This could potentially be verified by the future planned experimental
study of jet quenching for oxygen-oxygen  collisions at RHIC ($\sqrt{s}=0.2$ TeV)
\cite{CERN_STAR} and at the LHC
\cite{CERN_LHC,CERN_OOsum} ($\sqrt{s}=7$ TeV).
In light of this, it is highly desirable to perform a quantitative
analysis of jet quenching in O+O collisions at the RHIC and LHC energies
for scenarios with and without mQGP formation
in $pp$ collisions.

In the present work, by fixing free parameters
from fitting available data on $R_{AA}$ for heavy ion collisions, we perform
calculations of the nuclear modification
factor for 0.2 and 7 TeV O+O collisions for scenarios with and without
mQGP formation in $pp$ collisions  (we will often omit for clarity
adding ``in $pp$ collisions''), and for an intermediate
scenario when mQGP is formed only in $pp$ collisions
at the LHC energies. We use the jet quenching model of \cite{RAA08} (with some
improvements made in \cite{RAA20}), based  
on the LCPI approach \cite{LCPI1} to induced gluon
emission. We perform
calculations of the radiative energy loss
using the method suggested in \cite{RAA04}, which allows
to account for accurately the Coulomb effects in multiple parton
scattering.
An important feature of our calculations
is that we use a temperature dependent parametrization
of the running $\alpha_s$ (suggested in \cite{RAA20T}).
In \cite{RAA20T}
it was demonstrated that
the $T$-dependent QCD coupling largely solves the problem of difference
between the optimal values of $\alpha_s$ for the RHIC and LHC energies
\cite{RAA11,RAA20}. Use of the $T$-dependent QCD coupling allows  to avoid the
ambiguities for small systems ($pp$ and light ion collisions)
connected with the choice of $\alpha_s$ for large and small size QGP,
because the parameters, fitted to data for heavy ion collisions,
automatically fix $\alpha_s$ for small size QGP.
This improves the reliability of the results, as compared
to our previous analyses of jet quenching in $pp$ collisions
\cite{Z_pp_PRL,Z_pp13,Z_pp16}.
There we used a unique $\alpha_s$, which does not depend on the local
QGP temperature. The disadvantage of this ansatz is that it
leaves ambiguous how to choose $\alpha_s$ for the mQGP fireball
in $pp$ collisions.
Note that the issue of jet quenching in light ion collisions was addressed
recently in \cite{W_mQGP1}
within the BDMPS formalism \cite{BDMPS1} describing
multiple parton scattering in terms of the transport coefficient $\hat{q}$.
However, the scenario with mQGP formation was not considered in \cite{W_mQGP1}.

The plan of the paper is as follows.  In Sec. 2, we briefly review
the theoretical framework.  In  Sec. 3,  we discuss
the models of the QGP fireballs in $AA$ and  
in $pp$ collisions used in our jet quenching calculations. In Sec. 4 we first
perform the
$\chi^2$ fit  of  the  experimental data from the LHC
on $R_{AA}$ for Pb+Pb and Xe+Xe collisions for determining the optimal
free parameters. Then, we present results for $R_{pp}$ and $R_{AA}$ and for
the elliptic flow coefficient $v_2$
for heavy ion and for O+O collisions obtained with the optimal free parameters
for scenarios without and with mQGP formation. We conclude with
a brief discussion of jet quenching in $pA$ collisions.
Conclusions are contained in Sec. 5.

\section{Basic aspects of the jet quenching scheme}
Our  theoretical scheme for calculation of the medium modification
factors is similar
to  the  one  used  in  our  previous jet quenching  analyses
(see \cite{RAA08,Z_pp13,RAA20,RAA20T}).
For this reason, here we only outline its main aspects.
For $AA$ collisions we write the nuclear modification factor $R_{AA}^{st}$
for given impact parameter $b$
(we  consider  the  midrapidity  region  around $y=0$)

\beq
R_{AA}^{st}(b,\pb_{T},y)=\frac{{dN(A+A\rightarrow h+X)}/{d\pb_{T}dy}}
{T_{AA}(b){d\sigma(p+p\rightarrow h+X)}/{d\pb_{T}dy}}\,,
\label{eq:40}
\eeq
where $T_{AA}(b)=\int d\ro T_{A}(\ro) T_{A}(\ro-\bb)$ is the nuclear
profile function,
$T_{A}(\ro)=\int dz \rho_A(\sqrt{\rho^2+z^2})$ is the nuclear thickness function
(with $\rho_A$ the nuclear density).
The differential yield of the 
process $A+A\to h+X$
in the nominator on the righthand side of (\ref{eq:40}) reads
\beq
\frac{dN(A+A\rightarrow h+X)}{d\pb_{T} dy}=\int d\ro T_{A}(\ro+\bb/2)
T_{A}(\ro-\bb/2)
\frac{d\sigma_{m}(N+N\rightarrow h+X)}{d\pb_{T} dy}\,,\,\,\,
\label{eq:50}
\eeq
where $d\sigma_{m}/d\pb_{T} dy$ is the medium-modified cross section,
given by 
\beq
\frac{d\sigma_{m}(N+N\rightarrow h+X)}{d\pb_{T} dy}=
\sum_{i}\int_{0}^{1} \frac{dz}{z^{2}}
D_{h/i}^{m}(z, Q)
\frac{d\sigma(N+N\rightarrow i+X)}{d\pb_{T}^{i} dy}\,.\,\,\,
\label{eq:60}
\eeq
Here $\pb_{T}^{i}=\pb_{T}/z$ is the transverse momentum
of the initial hard parton, 
${d\sigma(N+N\rightarrow i+X)}/{d\pb_{T}^{i} dy}$ is the
ordinary pQCD hard cross section, and 
$D_{h/i}^{m}(z,Q)$ is the medium-modified FF
for transition of the initial hard parton $i$, with virtuality $Q\sim p^i_T$,
to the final hadron $h$. The FF $D_{h/i}^{m}$ 
 can symbolically be written as the triple $z$-convolution 
\beq
D_{h/i}^{m}(Q)\approx D_{h/j}(Q_{0})
\otimes D_{j/k}^{in}\otimes D_{k/i}^{DGLAP}(Q)\,,
\label{eq:70}
\eeq
where $D_{k/i}^{DGLAP}$ is the DGLAP parton FF for $i\to k$ transition,
$D_{j/k}^{in}$ describes the in-medium $j\to k$ transition
due to induced gluon emission, and 
$D_{h/j}$ is the ordinary vacuum FF for transition of the parton $j$ to
the final hadron $h$.
For ${D}_{h/j}(z,Q_0)$ we use the KKP
\cite{KKP} parametrization with $Q_0=2$ GeV.
We compute the DGLAP FFs
$D_{k/i}^{DGLAP}$
with the help of the PYTHIA event generator \cite{PYTHIA}.

The main ingredient in our analysis is the FF
$D_{j/k}^{in}$, which describes the medium effects.
We evaluate $D_{j/k}^{in}$ via the induced
gluon spectrum $dP/dx$ ($x$ is the gluon fractional momentum) in the
approximation of independent gluon
emission \cite{RAA_BDMS}.
A detailed description of the method for evaluation
of the induced gluon spectrum within the LCPI formalism
and the implementation of the approximation
of independent gluon emission for
calculation of  $D_{j/k}^{in}$ can be found in \cite{Z_phot,RAA20}.

The induced gluon spectrum $dP/dx$ depends on the QGP fireball
density/temperature profile along the jet trajectory.
Since the QGP temperature depends crucially
on the size of the collision system, 
for a joint analysis of jet modification
for large and small size collision systems it is 
important to account for the temperature dependence
of the in-medium QCD coupling.
In the present analysis we calculate the gluon induced spectrum
for running temperature dependent $\alpha_s$.
As in \cite{RAA20T},
we use the paramertization of $\alpha_s(Q,T)$ in the form 
\beq
\alpha_s(Q,T) = \begin{cases}
\dfrac{4\pi}{9\log(Q^2/\Lambda_{QCD}^2)}  & \mbox{if } Q > Q_{fr}(T)\;,\\
\alpha_{s}^{fr}(T) & \mbox{if }  Q_{fr}(T)\ge Q \ge cQ_{fr}(T)\;, \\
\alpha_{s}^{fr}(T)\times(Q/cQ_{fr}(T)) & \mbox{if }  Q < cQ_{fr}(T)\;, \\
\end{cases}
\label{eq:80}
\eeq
where
$Q_{fr}(T)=\Lambda_{QCD}\exp\left\lbrace
{2\pi}/{9\alpha_{s}^{fr}(T)}\right\rbrace$ (in the present analysis we
take $\Lambda_{QCD}=200$ MeV).
We perform calculations for $c=0.8$ and $c=0$.
We take $Q_{fr}=\kappa T$, and perform fit of the free parameter $\kappa$ using
data on the nuclear modification factor $R_{AA}$ for heavy ion collisions.
The parametrization (\ref{eq:80}) with nonzero $c$
is supported by the lattice results for the in-medium $\alpha_s$ in the
coordinate space via calculation of the free energy of a static heavy
quark-antiquark pair \cite{Bazavov_al1}.
The results of \cite{Bazavov_al1} give $\alpha_s(r,T)$ that becomes
close to the ordinary vacuum QCD coupling
$\alpha_s(Q)$ with $Q\sim 1/r$ at $r\ll 1/T$.
In the infrared region it reaches maximum at
$r\sim 1/\kappa T$ with $\kappa\sim 4$, and then it falls to
zero. With identification $r\sim 1/Q$, this behavior is
similar to that obtained in the momentum representation
within the functional renormalization group \cite{RG1}. 
The choice $c=0$ is similar to the model with $\alpha_s$
frozen in the infrared region \cite{Stevenson,DKT}.
Details of incorporating the $T$-dependence of $\alpha_s$
into the LCPI scheme can be found in \cite{RAA20T}.
It should be emphasized that the use of the running $\alpha_s$ is
very important in dealing with jet quenching simultaneously
in a large size QGP for heavy ion collisions and in
a smaller QGP for light ion and $pp$ collisions,
because the typical virtuality in induced gluon
emission increases with decreasing medium size.
The growth of the virtuality scale from heavy ion
to $pp$ collisions turns out to be quite large (by a factor of
 $\sim 2$ \cite{Z_pp13}).

We calculate the induced gluon spectrum
with the Debye mass obtained in the lattice analysis \cite{Bielefeld_Md}, 
and take $m_{q}=300$ and $m_{g}=400$ MeV 
for the light quark and gluon quasiparticle masses
in the QGP \cite{LH}. In numerical calculations for the $T$-dependence of
the Debye mass (and of $\alpha_s(Q,T)$) we define the temperature
via the lattice entropy density $s(\tau)$ obtained in \cite{t-lat}.
For a given fireball entropy, this procedure gives $T$ which
is somewhat larger than the ideal gas temperature.

For evaluation of $R_{AA}$ for a given centrality
range $\Delta c$, we perform accurate averaging over the geometry
of $AA$ collision and over the jet trajectories in the medium.
As in \cite{RAA08} we performed calculations
for a QGP fireball with a uniform density distribution in the transverse plane,
and with the Bjorken 1+1D longitudinal expansion \cite{Bjorken}.
The approximation of a uniform fireball is a crucial simplification,
because in this case the medium density profile along the fast parton
trajectory is independent of the jet production point (only its length $L$
varies). This approximation greatly reduces the computational time,
since we can first tabulate the $L$-dependence of
the induced gluon spectrum, and then to use it
in computations of the FFs for arbitrary geometry of the jet production.

We evaluate the medium modification factor $R_{pp}$ for
$pp$ collisions in a similar way. However, for $pp$ collisions
we need only the medium modification factor for the whole range
of the impact parameter.
The azimuthal dependence of jet quenching in mQGP is irrelevant
for calculation of $R_{pp}$.
Therefore, averaging over the geometry of the $pp$ collision
is simply reduced to
averaging of the medium modified FFs over the parton path length $L$
in the medium.
We perform averaging over $L$ for a symmetric fireball of radius $R_f$
produced in the central $pp$ collision.
This procedure seems to be reasonable, since the jet production should
be dominated by the head-on
collisions, and the azimuthal asymmetry is anyway irrelevant for $R_{pp}$.
We evaluate the $L$-distribution
using the MIT bag model quark density
(we assume that quarks and gluons have similar
distributions in the transverse coordinates).
The $L$-distribution turns out to be strongly peaked around $L\sim R_f$.
For this reason the $L$-fluctuations lead to a small change of $R_{pp}$
as compared to the $R_{pp}$ calculated for
$L=R_f$.

As in our previous jet quenching analyses,
we calculate the parton hard cross sections 
in  the LO pQCD with the CTEQ6 \cite{CTEQ6} parton distribution functions.
For nuclei we use the nuclear parton distributions
with the EPS09 correction \cite{EPS09}.
To simulate the higher order effects
we take for the virtuality scale in $\alpha_{s}$ the value 
$aQ$ with $a=0.265$ (as in the PYTHIA event generator \cite{PYTHIA}).
This gives for $pp$ collisions a fairly good description of the
$p_{T}$-dependence of the hadron spectra for the scenario without mQGP
formation. In principle,  for scenario 
with mQGP formation in $pp$ collisions
the perturbative cross section in formula (\ref{eq:20}) should
differ from that for the scenario without jet quenching
due to presence of $R_{pp}$ on the righthand side of (\ref{eq:30}).
For a fully consistent treatment of $R_{pp}$
one should use a bootstrap procedure
with adjusting the free parameters
to fit the experimental $pp$ cross section.
However,  this consistent procedure does not make much sense (at least at this stage).
The point is that the hadron spectra 
have a very steep $p_{T}$-dependence (as compared to 
a weak $p_{T}$-dependence of $R_{pp}$).
For this reason $R_{pp}$ (and $R_{AA}$) is only weakly sensitive
to possible difference in the perturbative cross sections
for scenarios  without and with jet quenching in $pp$ collisions.
To  roughly mimic the change of the pQCD $pp$ cross section in scenario
with mQGP, we performed calculations of $R_{pp}$ and $R_{AA}$
using for the $\alpha_s$ virtuality scale in the hard cross sections 
$a'Q$ with $a'=0.13$ for the LHC energies and $a'=0.205$ for
the RHIC energy $\sqrt{s}=0.2$ TeV. These values  of $a'$ give
the $p_T$-dependence of the righthand side of (\ref{eq:20})
which is reasonably close to that for perturbative cross section
without jet quenching for the hard scale $aQ$ with the PYTHIA prescription
$a=0.265$ (which we use
for scenario without mQGP formation).  However, we have found
that the effect of the modification of the hard cross section
for the mQGP version on our predictions for $R_{pp}$ and $R_{AA}$ is very
small, as expected.

As in \cite{RAA08}, we treat the effect of the
collisional energy loss, which is
relatively small \cite{BSZ,Z_coll,Gale_coll}, as a small perturbation
to the radiative jet quenching. 
We incorporate it in calculating the FFs $D_{j/k}^{in}$ 
by renormalizing the initial QGP temperature 
according to the relation
\beq
\Delta E_{rad}(T^{\,'}_{0})=\Delta E_{rad}(T_{0})+\Delta E_{col}(T_{0})\,,
\label{eq:90}
\eeq
where
$\Delta E_{rad/col}$ is the radiative/collisional energy loss, $T_{0}$
is the true initial QGP temperature,
and $T_0^{\,'}$ is the new temperature. 

We conclude this section with a discussion of the possible errors
due to the neglect of the transverse QGP motion and the use of a flat
initial entropy density. We begin with the errors due to the neglect of
the QGP transverse flow. It was shown long ago by Baier, Mueller and
Schiff \cite{BMS_flow}, within the BDMPS formalism \cite{BDMPS1}, that the
flow corrections to predictions obtained using the Bjorken model should
be small. This occurs due to an almost complete compensation between the
enhancement  of parton energy loss caused by increase of the
fireball size and its suppression caused by reduction of the transport
coefficient due to the QGP expansion. 
More recently, a detailed numerical analysis of the flow effects in
jet quenching at the RHIC and LHC energies, based on the pQCD ansatz for the
radiative parton energy loss, performed by Betz and Gyulassy \cite{BG_flow}
confirmed that the effect of the QGP transverse expansion on the medium
suppression of high-$p_T$ spectra is small.
In \cite{BG_flow} it was found that for 0.2 TeV Au+Au and 2.76 TeV Pb+Pb
collisions the change of
$R_{AA}$ due to the transverse flow is $\lsim 1-2$\% (see Fig. 2 of
\cite{BG_flow}),
and it holds both for central and noncentral collisions.
The results of \cite{BG_flow} show that for heavy ion collisions
the errors of our predictions due to the neglect of the QGP transverse
expansion should be small. 
It is important that for noncentral 0.2 TeV Au+Au collisions for $20-30$\%
centrality bin the authors of \cite{BG_flow} obtained very small flow
corrections to $R_{AA}$ both for the jet azimuthal angle
$\phi\sim 0$ and $\phi\sim \pi/2$ (i.e., for the jet directions
along the short and long fireball semi-diameters).
For Au+Au collisions in $20-30$\% centrality bin the short fireball
semi-diameter is $\sim 2.5$ fm. Since this size is close to the oxygen radius,
one can conclude that the flow correction to $R_{AA}$ in O+O collisions
should be small.
For the mQGP produced in $pp$ collisions
the fireball radius (see Eq. (\ref{eq:130}) below) is smaller than
the short fireball semi-diameter for Au+Au collisions for
$20-30$\% centrality by just $\sim 40$\%. Due to the smaller QGP temperature
(see Eq. (\ref{eq:140}) below) the QGP sound velocity for $pp$ collisions
should be smaller than that for Au+Au collisions.
This can reduce somewhat the transverse QGP velocity (which is approximately
proportional to the QGP squared sound velocity \cite{Ollit}) for $pp$
collisions. The above facts show that the magnitudes of the QGP
transverse flow for 0.2 TeV Au+Au collisions for $20-30$\%
centrality bin and $pp$ collisions should not differ drastically. 
Therefore, there is no reason to expect that the flow correction to $R_{pp}$
can be dramatically larger than that found in \cite{BG_flow} for $R_{AA}$
in noncentral Au+Au collisions. Thus, one can conclude that the errors
of our predictions due to the use of the Bjorken model should be quite
small. 

The situation with the errors due to the use a flat initial
entropy distribution in some sense is similar to that for the neglect of
the transverse flow. In this case one can
also expect a strong compensation between the enhancement  of the parton
energy loss in the regions where the density is overestimated
and its suppression in the regions where the density is underestimated.
The jet configurations that can potentially lead to the incomplete
compensation are the jet production near the surface of the fireball,
when $R_{AA}\sim 1$ for the jets heading outward, and $R_{AA}\ll 1$ for the
ones  heading to the core of the fireball. This nonlinear effect is possible
for heavy ion collisions, but cannot be important for small systems.
In general, it is clear that for small systems the sensitivity
of jet quenching to the form of the fireball density should
be weak, since for such systems we have a situation where
the typical formation length for induced gluon emission is of 
the order of the fireball size or larger.
In such a regime the form of the density profile along the fast parton
trajectory is not very important, and the parton energy loss is mostly
controlled by the total amount of the matter traversed by fast parton.
As far as the possible errors in $R_{AA}$ for heavy ion collisions are
concerned, even if they are several times larger than the flow corrections found in \cite{BG_flow}, they can lead only to a
relatively small change of the parameters fitted from heavy ion data.
For instance, for $R_{AA}\sim 0.2$ a change of $R_{AA}$ by $\sim 20$\%
can be mimicked by changing $\alpha_s$ by $\sim 3$\%
\footnote{This estimate can be easily obtained from the analytical relation
  between the energy loss and the medium suppression factor derived
  in \cite{RAA_BDMS}, considering that the radiative energy loss
  is dominated by the $N=1$ rescattering term (which is
  $\propto \alpha_s^3$).}.
Note that the necessary change of $\alpha_s$ to compensate
the errors $\sim 20$\% in $R_{AA}$ for heavy ion collisions will translate
to the errors $\lsim 10$\% in $R_{AA,pp}-1$ for small system.
Thus, it is clear that the approximation of a flat QGP density
cannot spoil significantly the quality of the extrapolation of
the jet quenching predictions from heavy ion collisions to collisions of small
systems. It is worth noting that,  in principle, even for an accurate
hydrodynamic fireball evolution it is impossible to avoid the errors of
the order of $\sim 10$\% in $R_{AA,pp}-1$ in extrapolating the theoretical
predictions from heavy ion collisions to small system collisions,
because there are also other considerable theoretical uncertainties in
jet quenching calculations (e.g. related to the treatment of multiple
gluon emission and to the choice of the number density of the color
centers, which is the basic input to calculation of the induced gluon spectrum).


\section{Models of the QGP fireballs for $AA$ and $pp$ collisions }
We perform calculations for simple Bjorken's 1+1D expansion
\cite{Bjorken} of the QGP. It gives the entropy density
$s(\tau)/s(\tau_0)=\tau_0/\tau$, where $\tau_0$ is the thermalization
time. At $\tau<\tau_{0}$ we use a linear parametrization
$s(\tau)=s(\tau_0)\tau/\tau_0$.
We set $\tau_{0}=0.5$ fm both for $AA$ and $pp$ collisions. But to understand
sensitivity of the results to $\tau_0$, we also perform calculations
for $\tau_0=0.8$ fm.
As in our previous jet quenching analyses, we model QGP by a 
system of the static Debye  screened color centers \cite{GW}.
We use the ideal gas model to relate the number density of the
QGP constituents to the entropy density. 
This gives the effective number density of the triplet color
centers $n(z)=bT^{3}(z)$ with
$b=9\xi(3)(N_f+4)/\pi^2\approx 7.125$ (for $N_f=2.5$).
As was said above,
we use the fireball model
with a uniform density distribution in the transverse plane.

\subsection{The QGP fireball in $AA$ collisions}
Let us start with $AA$ collisions.
To fix the average initial QGP entropy density for $AA$ collisions
we use the relation \cite{Bjorken}
\beq
s_{0}=\frac{C}{\tau_{0} S_{f}}\frac{dN_{ch}(AA)}{d\eta}\,,
\label{eq:100}
\eeq
where $S_{f}$ is the area of the overlap
region of two colliding nuclei,
and $C=dS/dy{\Big/}dN_{ch}(AA)/d\eta\approx 7.67$ \cite{BM-entropy} 
is the entropy/multiplicity ratio. 
We calculate $dN_{ch}(AA)/d\eta$ in the Glauber wounded  nucleon model
\cite{KN-Glauber} with parameters of the model as in our Monte-Carlo
Glauber analyses \cite{Z_MC1,Z_MC2},
which describe very well data on the midrapidity $dN_{ch}/d\eta$
in $0.2$ TeV Au+Au \cite{STAR02_Nch},
$2.76$ \cite{ALICE276_Nch} and $5.02$ TeV \cite{ALICE502_Nch} Pb+Pb, 
and $5.44$ TeV Xe+Xe \cite{ALICE544_Nch} collisions.
For heavy nuclei we use the Woods-Saxon nuclear density
$\rho_{A}(r)=\rho_{0}/[1+\exp((r-R_{A})/d)]$.
For Pb nucleus we take the parameters
$R_A=6.62$ and $d=0.546$ fm, as in the PHOBOS
Glauber model \cite{PHOBOS}, and 
for Xe nucleus we take
$d=0.54$ fm and  $R_{A}=(1.12A^{1/3}-0.86/A^{1/3})$ fm,
as in the GLISSANDO Glauber model \cite{GLISS2} (it 
gives $R_A\approx 5.49$ fm).
For our basic version we define $S_f$ for overlap of two circles with radius
$R=R_{A}+k d$ with $k=2$ ($R_{A}$ and $d$
are the parameters of the Woods-Saxon nuclear density).
To understand the sensitivity of the results to the choice of $k$, we also
performed calculations for $k=3$.
  The range $2\lsim k\lsim 3$
  seems to be reasonable for determining the fireball size, because,
  on the one hand,
  it guarantees that the fraction of the lost QGP volume
  is small. On the other hand,
  for such values of $k$ our ansatz with a flat fireball density
  does not lead to a considerable redistribution of the QGP density into
  regions where the true QGP density should vanish.
  Our calculations give a small variation of $R_{AA}$ in heavy ion
  collisions with $k$ in the range $2\lsim k\lsim 3$ (see below).

For O+O collisions we performed calculations for the harmonic
oscillator shell model nuclear density   
\bea
\rho_A(r)=\frac{4}{\pi^{3/2}r_A^3}\left[1+
  \frac{A-4}{6}\left(\frac{r}{r_A}\right)^2\right]\exp\left(r^2/r_A^2\right)\,,
\nonumber\\
r_A^2=\left(\frac{5}{2}-\frac{4}{A}\right)^{-1}
\left(\langle r_{ch}^2\rangle_A-\langle r_{ch}^2\rangle_p\right)\,
\label{eq:110}
\eea
with $\langle r_{ch}^2\rangle_A=7.29$ fm$^2$ for $^{16}$O and 
$\langle r_{ch}^2\rangle_p=0.7714$ fm$^2$ \cite{GLISS2}.
Also, to understand the sensitivity of the results to the $^{16}$O nuclear
density we have performed calculations for the Woods-Saxon
density.
For the Woods-Saxon
oxygen nuclear density we take $d=0.513$ fm \cite{ATDATA}
and $R_A=2.2$ fm (fixed through the condition of equal
$\langle r_{ch}^2\rangle$ for the Woods-Saxon and
the harmonic oscillator shell model nuclear density (\ref{eq:110})).
Our calculations show that the difference between $R_{AA}$ for the
harmonic oscillator
shell model and for the Woods-Saxon one is very small for the whole
range of centrality. We therefore present results only for
the oscillator shell model nuclear density.
Similarly to heavy ion collisions, for O+O collisions we 
define $S_f$ for overlap of two circles with radius
$R=R_{A}+kd$ and $k=2$, where $R_{A}$ and $d$ the parameters of the
Woods-Saxon nuclear density
(both for the harmonic oscillator shell model density and the
Woods-Saxon one). For heavy ions the results are not very sensitive to
variation of the nuclear geometry parameter $k$. But for light nuclei the
variation
of $R_{AA}$ with $k$ may be more noticeable, since the ratio
of the nuclear surface thickness to the nuclear radius becomes larger.
To illustrate dependence of the result on $k$, for O+O collisions
we also present results for $k=3$.

Fig. 1 shows the centrality dependence of the initial QGP temperature
obtained within the Glauber model with the help of the relation (\ref{eq:100})
for heavy ion and O+O collisions. 
We show the results for the ideal gas model, when
$s(T)=aT^3$ with $a=\frac{4\pi^2}{15}\left(8/3+7N_f/4\right)$
($a\approx 18.53$, if one takes $N_f=2.5$).
The QGP temperature calculated via (\ref{eq:100}) for the lattice entropy density
\cite{t-lat} is somewhat bigger than that  for the ideal gas model
(by $\sim 5-10$\% at $T\sim 300-450$, and by $\sim 15-25$\% at
$T\sim 150-200$ MeV).
From Fig. 1 one can see that for $7$ TeV O+O collisions
the initial fireball temperature is qualitatively similar to that
for $0.2$ TeV Au+Au collisions. We therefore expect
the QGP formation and jet quenching in 7 TeV O+O collisions.
However, for 0.2 TeV O+O collisions the initial temperature
for centrality $\lsim 30$\% is comparable to that expected for $pp$
collisions (see below). For this reason, there is some doubt whether
QGP formation in 0.2 TeV O+O collisions is possible, if there is no mQGP
formation in $pp$ collisions. But one should bear in mind
that O+O collisions have better conditions for collectivity in
the final state, because in this case the Knudsen number
becomes bigger by a factor of 2 as compared to that
for $pp$ collisions. In the present analysis we assume
that in 0.2 TeV O+O collisions a QGP is formed.

For a flat entropy distribution the model with the almond shaped
fireball with two cups
somewhat overestimates the anisotropy coefficient $\epsilon_2$, as compared
to that calculated with an accurate $b$-dependent entropy density
\cite{RAA20}.
This fact is immaterial for $R_{AA}$, which is practically insensitive
to azimuthal asymmetry of the fireball. 
But for calculation of the flow coefficient $v_2$
this model is too crude.
In order to have in our calculations
the fireball ellipticity which coincides with that for
the accurate Glauber model fireball density, as in \cite{RAA20},
we transform the almond shaped region
into an elliptic one (of the same area).
We perform calculations of $v_2$ for two choices of the fireball
eccentricity $\epsilon_2$. For the first variant, we calculate
$\epsilon_2$ in the optical Glauber wounded nucleon model,
and for the second variant we use $\epsilon_2$ obtained in
the  Monte-Carlo Glauber model of \cite{Z_MC1,Z_MC2}.
The Monte-Carlo version gives $\epsilon_2$, which, contrary to the
optical model one, does not vanish for central collisions
(due to density fluctuations).
But for noncentral collisions of heavy nuclei predictions of the optical
and Monte-Carlo Glauber models for
$\epsilon_2$ do not differ significantly.
Our treatment of the flow coefficient $v_2$ for the Monte-Carlo
version of $\epsilon_2$ ignores possible decorrelation between the
participant plane and the true reaction plane. The 
effect of this decorrelation should be small for heavy nuclei
(see discussion in \cite{RAA20}).
However, for light ion collisions, such as O+O in our case,
when the density fluctuations become large, our calculations
of $v_2$ become clearly less robust. It worth noting that
for light nuclei even the initial anisotropy coefficient $\epsilon_2$
itself cannot be calculated accurately in the Monte-Carlo
Glauber model, because the theoretical predictions
become sensitive to the size of the entropy sources in the whole range
of centrality. For this reason, our results for $v_2$ in O+O
collisions are of illustrative character. But nevertheless
they allow to understand the relation between the initial fireball
ellipticity $\epsilon_2$ and the flow coefficient $v_2$ for light
nucleus collisions. We perform calculation of $v_2$ using $\epsilon_2$ for
the Gaussian sources with width $0.7$ fm.
In Fig. 2 we show the theoretical prediction for $\epsilon_2$
for 0.2 and 7 TeV O+O collisions,
obtained in the optical Glauber wounded nucleon model and in the
Monte-Carlo one, which we use in our calculations of $v_2$.
To illustrate the sensitivity of $\epsilon_2$
to the source width in the Monte-Carlo scheme we also plot in
Fig.~2 predictions for the Gaussian width $0.4$ fm.
From Fig. 2 one sees that there is a significant difference between the
results for the optical and Monte-Carlo versions, and that the Monte-Carlo
model $\epsilon_2$ is rather sensitive to the entropy source size.
The curves for $\epsilon_2$ for heavy ion collisions can be found in
\cite{RAA20}.
\begin{figure} 
\begin{center}
\includegraphics[height=5.5cm]{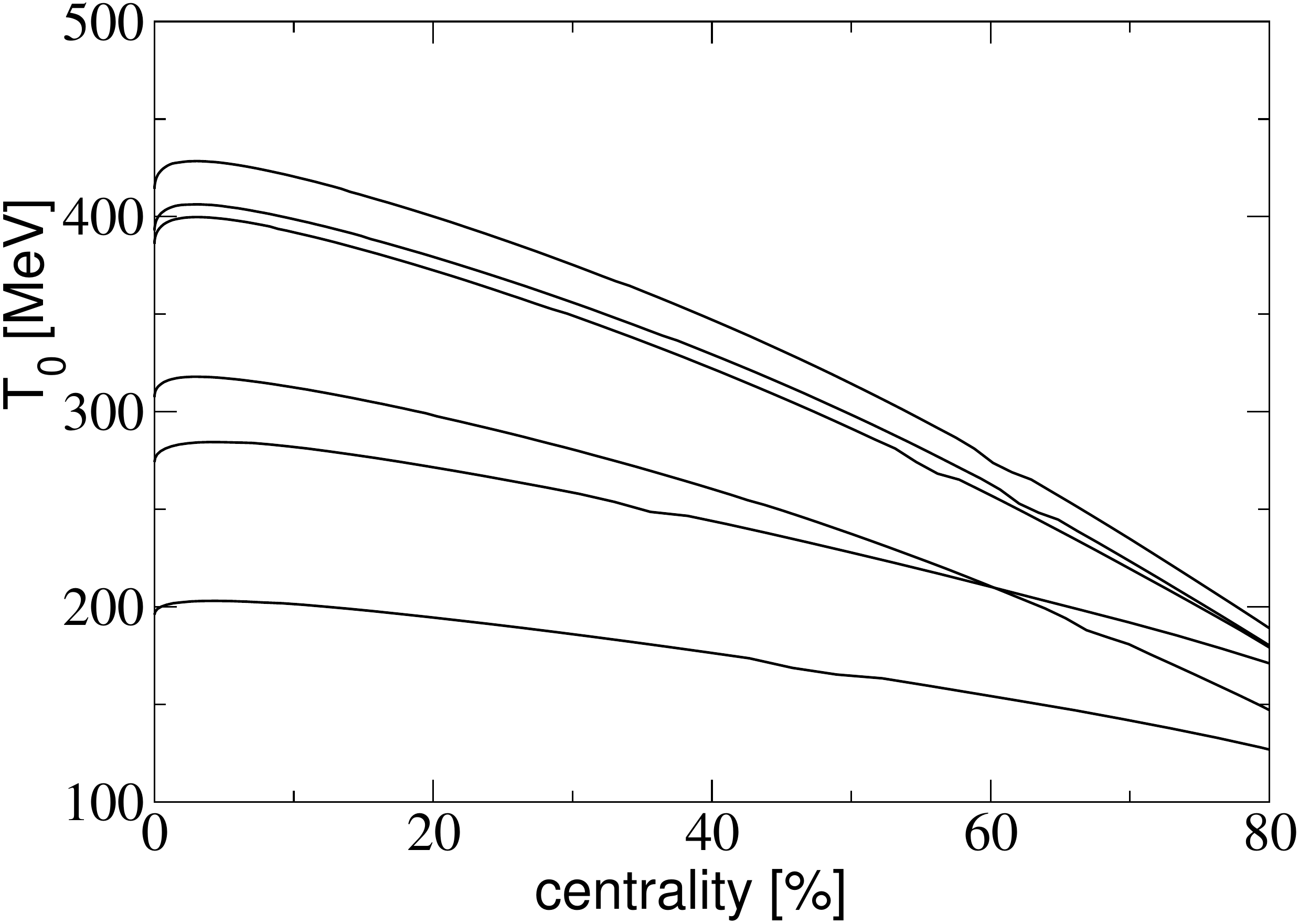}  
\end{center}
\caption[.]{Centrality dependence of the initial fireball temperature
at  $\tau_0=0.5$ fm for the ideal gas model obtained
in the Glauber model via the average entropy
density for (from top to bottom at low centrality):
$5.02$ and $2.76$ TeV Pb+Pb,  
  $5.44$ TeV  Xe+Xe, $0.2$ TeV Au+Au,
$7$ and $0.2$ TeV O+O
collisions.
 }
\end{figure}
\begin{figure} 
\begin{center}
\includegraphics[height=5.5cm]{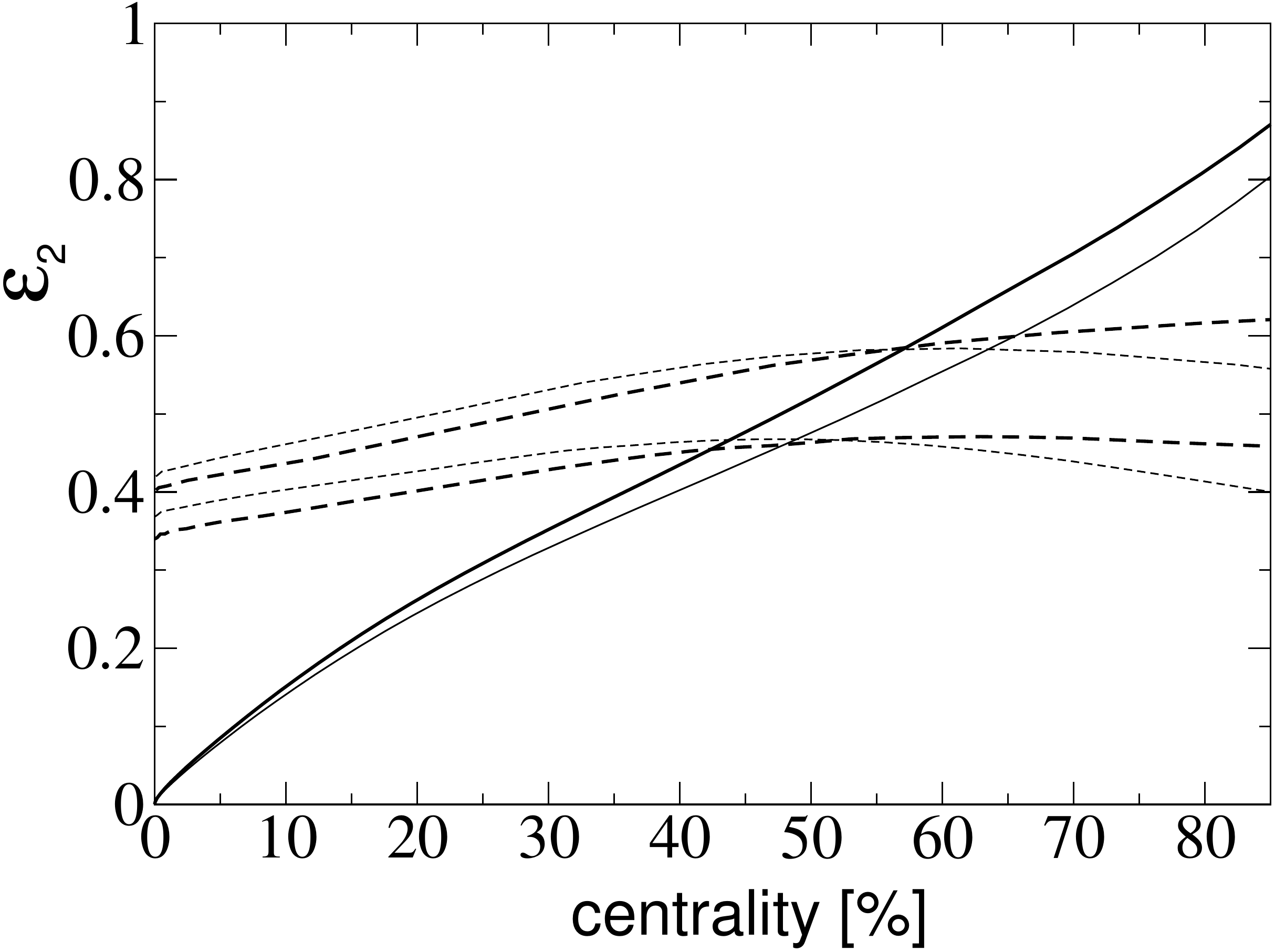}  
\end{center}
\caption[.]{Centrality dependence of the initial fireball
  eccentricity $\epsilon_2$ for O+O collisions
  at $\sqrt{s}=0.2$ TeV (thin lines) and $7$ TeV (thick lines)
  calculated in the optical 
  (solid) and in the Monte-Carlo (dashed) Glauber wounded nucleon model.
  The curves for the Monte-Carlo version are obtained for
  the Gaussian width parameter of the entropy sources (top
  to bottom) $0.4$ and $0.7$ fm.
 }
\end{figure}

For $AA$ collisions we use the medium life/freeze-out
time
$\tau_{f.o.}\approx 1.05\times(dN_{ch}/d\eta)^{1/3}$, which is
supported by the pion interferometry at RHIC \cite{STAR_HBT} and
LHC \cite{ALICE_HBT}.
However, the exact value of the QGP fireball life time is not very
crucial to our analysis, since the jet quenching effect of the final stage with
$\tau\sim\tau_{f.o.}$ is very small \cite{RAA08,RAA20}.
Note that, from the viewpoint of jet quenching,
the nature of the medium in the final stage, whether it is QGP or hadron gas,
is not vary crucial, because for a given entropy the energy loss
in QGP and hadronic gas turn out to be close to each other
\cite{Baier_q}.

\subsection{The mQGP fireball parameters for $pp$ collisions}
We now come to the mQGP fireball in $pp$ collisions.
Our treatment of the mQGP fireball is similar to that of
Ref. \cite{Z_pp13}.
To fix $T_{0}$,  we use the relation
(\ref{eq:100}) with  $dN_{ch}(AA)/d\eta$ replaced by the UE multiplicity
density in $pp$ collisions $dN_{ch}^{ue}(pp)/d\eta$.
We take $S_f=\pi R_f^2$, where $R_f$ is the effective radius of the mQGP
fireball in the $pp$ collision (effective in the sense that it
is an average radius for the whole range of the impact parameter).
We determine $R_f$ via the UE multiplicity density
$dN_{ch}/d\eta$ using the prediction for $R_{f}$ obtained 
in numerical simulations performed in \cite{glasma_pp} 
within the IP-Glasma model \cite{IPG12}.
We use the parametrization of $R_{f}$ from \cite{glasma_pp} 
as a function of $dN_{g}/dy$
given in \cite{RPP}
\beq
 R_{f} = 1\,{\rm fm}\times f_{pp}\left(\sqrt[3]{dN_g/dy}\right)\,,
\nonumber
\eeq
\beq
 f_{pp}(x) = 
\left\{ \begin{array}{ll}
         0.387 + 0.0335 x + 0.274\,x^2 - 0.0542\,x^3  & \mbox{if $x < 3.4$,}\\
         1.538 & \mbox{if $x \geq 3.4 $.}
\end{array} \right.
\label{eq:120} 
\eeq
We take $dN_{g}/dy=bdN_{ch}/d\eta$
with $b=C45/2\pi^{4}\xi(3)\approx 2.13$.
For $pp$ collisions at $\sqrt{s}=0.2$ we calculate the UE multiplicity density
using the $K_{ue}$ from PHENIX measurement
\cite{PHENIX_dA} and the minimum bias 
non-single-diffractive event multiplicity density $dN_{ch}^{mb}/d\eta=2.65$ from
the UA1 Collaboration \cite{UA1_pp}. In the plateau region
this gives $dN_{ch}^{ue}/d\eta\approx 5.79$.
We evaluate the UE multiplicity  
at $\sqrt{s}=2.76$ and $5.02$ TeV by interpolating the ATLAS 
data \cite{ATLAS_UE_Nch} between $\sqrt{s}=0.9$ and $7$ TeV
that give in the plateau region $dN_{ch}^{ue}/d\eta\approx 7.5$ and $13.9$.
Assuming that $dN_{ch}^{ue}/d\eta\propto s^{\delta}$,
we obtained $dN_{ch}^{ue}/d\eta\approx 10.5(12.6)$ for $\sqrt{s}=2.76(5.02)$ TeV.
Using these values of
$dN_{ch}^{ue}/d\eta$ we obtain
\beq
R_{f}[\sqrt{s}=0.2,2.76,5.02,7\,\, \mbox{TeV}]
\approx[1.26,1.44,1.49,1.51]\,\,\mbox{fm}\,.
\label{eq:130}
\eeq
Then, using the above values of
$dN_{ch}^{ue}/d\eta$ and $R_f$ from (\ref{eq:130}),
we obtain for the initial temperature of the mQGP fireball
for the ideal gas entropy and for the lattice entropy \cite{t-lat}
(numbers in brackets)
\beq
T_{0}[\sqrt{s}=0.2,2.76,5.02,7\,\,\mbox{TeV}]
\approx[195(226),217(247),226(256),232(261)]\,\,\mbox{MeV}\,.
\label{eq:140}
\eeq
As one can see,  for the lattice entropy $T_0$
is larger by $\sim 10-15$\% than $T_0$ determined
in the ideal gas model.

\section{ Numerical results}

\subsection{Optimal $\kappa$ from the $\chi^2$ fit}
We have adjusted the free parameter $\kappa$ by $\chi^2$ fitting
of the LHC data on $R_{AA}$ for heavy ions collisions for
centrality $\lsim 30$\% and the transverse momentum
$10<p_{T}<120$ GeV.
We have used data from ALICE \cite{ALICE_r276}, ATLAS \cite{ATLAS_r276},
and CMS \cite{CMS_r276} for $2.76$ TeV Pb+Pb collisions; 
from ALICE \cite{ALICE_r502}, ATLAS \cite{ATLAS_r502},
and CMS \cite{CMS_r502} for $5.02$ TeV Pb+Pb collisions; 
and for $5.44$ TeV Xe+Xe collisions
from ALICE \cite{ALICE_r544}, ATLAS \cite{ATLAS_r544},
and CMS \cite{CMS_r544}. 
As usual, we define $\chi^2$ as
\beq
\chi^2 = \sum_i^N \frac{(f_i^{exp} - f_i^{th})^2}
    {\sigma_{i}^2}\,,
    \label{eq:150}
    \eeq
    where $N$ is the number of the data points,
and $\sigma_i^2=\sigma_{i,stat}^2+\sigma_{i,sys}^2$.
We calculate $\chi^2$ using 
the theoretical $R_{AA}$ obtained with the help of a cubic spline
interpolation in $\kappa$ using a pre-prepared grid with steps
$\Delta \kappa/\kappa \sim 0.1$.
The optimal values of $\kappa$, and the corresponding
$\chi^2/d.p.$ ($\chi^2$ per data point), for the versions with
and without mQGP formation in $pp$ collisions,
obtained using the geometrical parameter $k=2$ and the parametrization (\ref{eq:80}) with $c=0.8$,
are summarized in Table I.
In Table I we present results
for each energy (and process) separately  and for the combined fit for
all LHC data.
We show for $\kappa$ the standard errors
(i.e. corresponding to $\Delta \chi^2=1$), and the 95\% confidence
intervals (CIs)
for the 95\% quantile of the $\chi^2$-distribution.
From Table I it is seen that for all the cases we have a quite good
fit quality ($\chi^2/d.p.\lsim 1$).
The optimal values of $\kappa$ for different energies/processes
are very similar.
We find, as one could expect, that scenario
with mQGP formation requires somewhat smaller values of $\kappa$
(the reduction of $\kappa$ for this version is $\sim 1$).
We have also performed fitting for the geometrical parameter
  $k=3$.
In this case, for 
the optimal $\kappa$ for all the LHC data we obtained
$\kappa \approx 3.74(2.98)$ for scenarios
without(with) mQGP formation with
a quite good fit quality ($\chi^2/d.p.\sim 0.7$).
Our fit for $\tau_0=0.8$ fm gives the values of $\kappa$ that are smaller
by $\sim 0.22(0.12)$ for scenarios without(with) mQGP formation.
For calculations with $c=0$ in (\ref{eq:80}), the optimal values of $\kappa$  are larger
than that for $c=0.8$ by $\sim 0.4(0.05)$ for versions without(with)
mQGP formation. 
 
\begin{table}
  \hspace{-.2cm}
         {
           \footnotesize
  \begin{tabular}{c|ccc|ccc}
    \hline
& \multicolumn{3}{c}{Without mQGP} &
\multicolumn{3}{|c}{With mQGP} \\
\cline{2-7}
& $\kappa$ &95\% CI  &$\chi^2/d.p.$  & $\kappa$ &
95\% CI &  $\chi^2/d.p.$  \\
\hline
Pb+Pb~2.76~TeV  &$3.44_{-0.04}^{+0.04}$
&$(3.16,3.7)$ & $0.75$ &
$2.57_{-0.05}^{+0.05}$& $(2.3,2.82)$ & $1$\\
\hline
Pb+Pb~5.02~TeV  &$3.47_{-0.06}^{+0.06}$&$(3.02,3.82)$ & $0.76$ &
$2.5_{-0.08}^{+0.08}$&$(2.1,2.8)$ & $1$\\
\hline
Xe+Xe~5.44~TeV  &$3.59_{-0.08}^{+0.08}$&$(2.92,4.19)$&$0.47$ &
$2.52_{-0.12}^{+0.109}$&$(2,3.3)$ & $0.51$\\
\hline
All LHC data &$3.47_{-0.03}^{+0.03}$&$(3.13,0.3.78)$ & $0.68$ &
$2.55_{-0.04}^{+0.04}$&$(2.16,2.88)$ & $0.87$\\
\hline
  \end{tabular}
}  
  \caption{
    Optimal values of $\kappa$ with $1\sigma$ standard errors,
    95\% CI and corresponding $\chi^2/d.p.$,
    obtained for $k=2$, $c=0.8$ at $\tau_0=0.5$ fm,
    $\tau_{max}=1.05\times(dN_{ch}/d\eta)^{1/3}$
    for $10<p_T<120$ GeV 
    for different energies/processes separately and for all LHC data
    for scenarios without (left)
    and with (right) mQGP formation in $pp$ collisions.}
\label{chi2fit}
\end{table}

\subsection{Predictions for $R_{pp}$, $R_{AA}$ and azimuthal
asymmetry $v_2$}  
\begin{figure} [!h]
\begin{center}
\includegraphics[height=4.5cm]{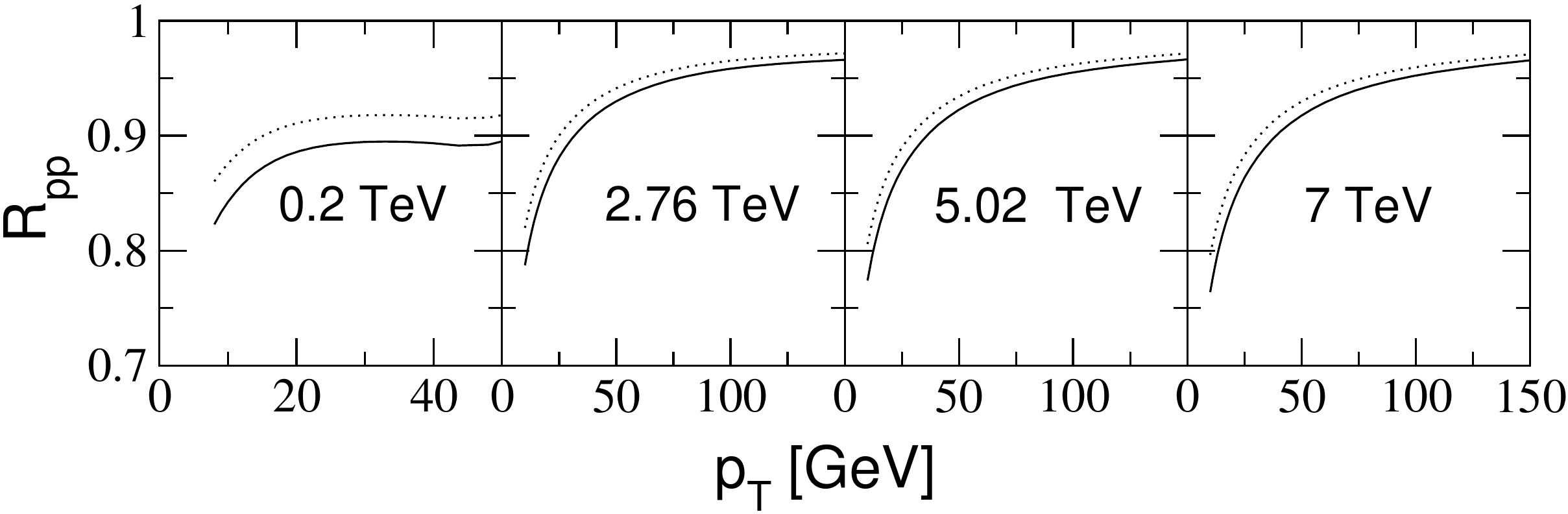}  
\end{center}
\caption[.]
{
  $R_{pp}$ of charged hadrons for $0.2$, $2.76$, $5.02$ and $7$ TeV 
  $pp$ collisions
  from our calculations
  for $\tau_0=0.5$ (solid) and $0.8$ fm (dotted)
  with $\kappa=2.55$ obtained by fitting all the LHC data on $R_{AA}$ for
  heavy ion collisions with the geometrical parameter $k=2$. 
}
\end{figure}
\begin{figure} [!h]
\begin{center}
\includegraphics[height=8.5cm]{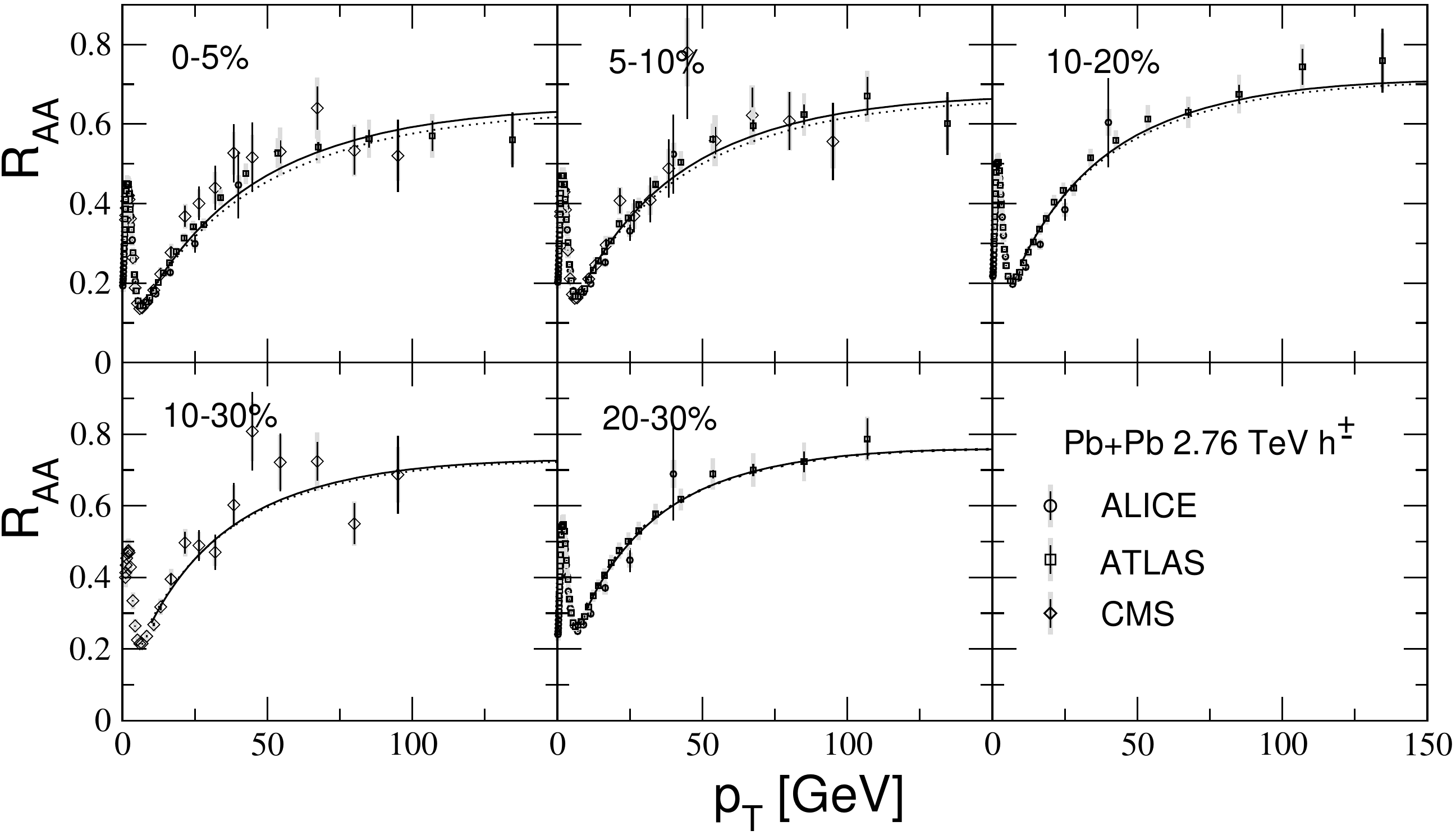}  
\end{center}
\caption[.]
{
$R_{AA}$ of charged hadrons for $2.76$ TeV Pb+Pb collisions
  from our calculations with $k=2$, $c=0.8$, $\tau_0=0.5$ fm for scenarios
  without (solid) and with (dotted)
  mQGP formation in $pp$ collisions for
  the optimal parameters $\kappa=3.44$ and $2.57$
obtained by fitting $R_{AA}$ in the range $10<p_T<120$ GeV.  
  Data points are from ALICE \cite{ALICE_r276}, ATLAS \cite{ATLAS_r276},
  and CMS \cite{CMS_r276}.
}
\end{figure}
\begin{figure} [!h]
\begin{center}
\includegraphics[height=8.5cm]{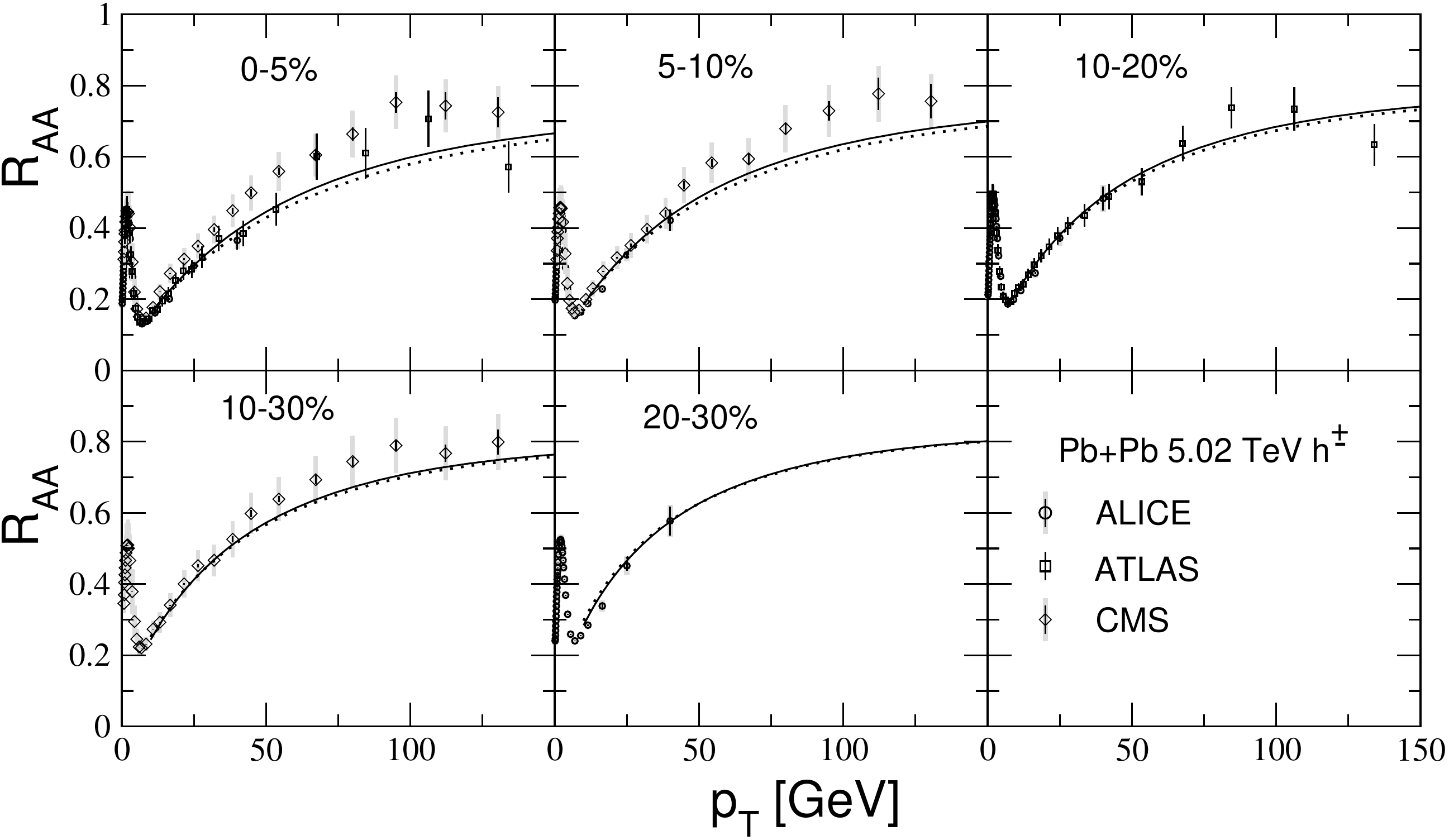}  
\end{center}
\caption[.]
        {Same as in Fig.~4 for $\sqrt{s}=5.02$ TeV for the
          optimal parameters $\kappa=3.47$ and $2.5$.
  Data points are from ALICE \cite{ALICE_r502}, ATLAS \cite{ATLAS_r502},
  and CMS \cite{CMS_r502}.
}
\end{figure}
\begin{figure}[!h]
\begin{center}
\includegraphics[height=8.5cm]{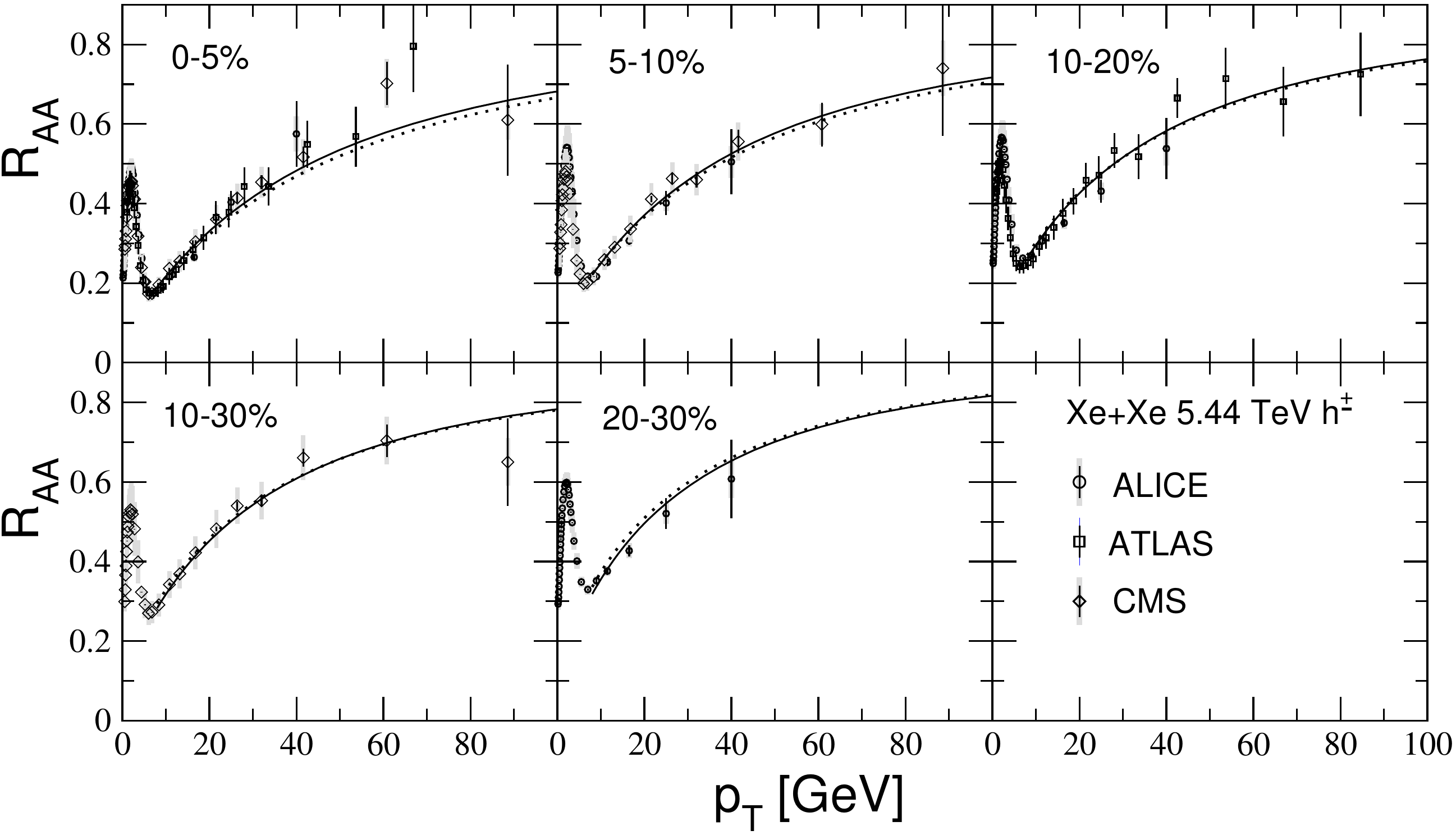}  
\end{center}
\caption[.]
        {
Same as in Fig.~4 for $5.44$ TeV Xe+Xe collisions for the
optimal parameters $\kappa=3.59$ and $2.52$.
          Data points are from ALICE \cite{ALICE_r544}, ATLAS \cite{ATLAS_r544},
          and CMS \cite{CMS_r544}.
}
\end{figure}
\begin{figure}[!h]
\begin{center}
\includegraphics[height=8.5cm]{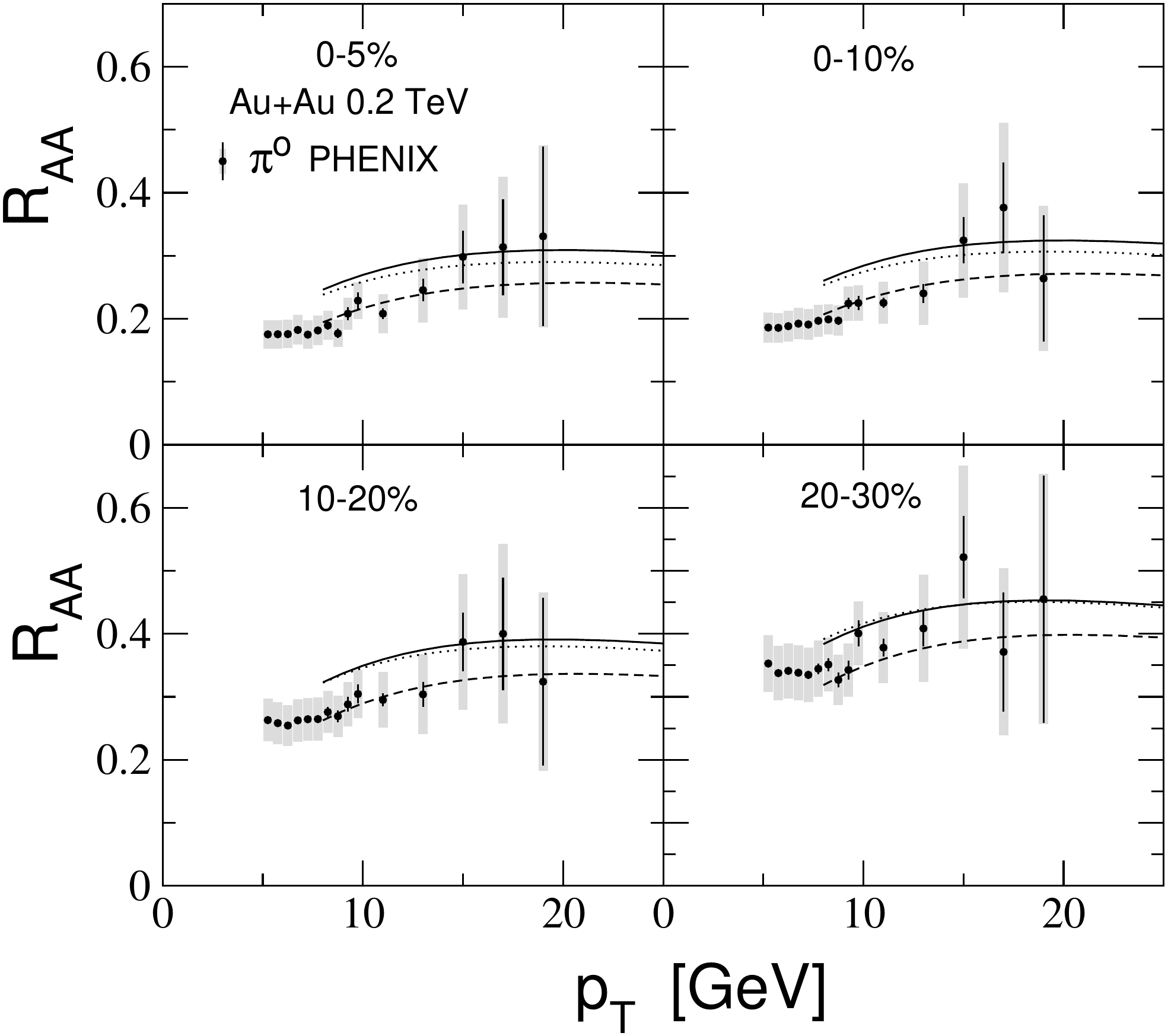}  
\end{center}
\caption[.]{Same as in Fig.~4 for $R_{AA}$ of $\pi^{0}$ in
  $0.2$ TeV  Au+Au collisions for the optimal parameters
  $\kappa=3.47$ and $2.55$ determined by fitting all the LHC
  data on $R_{AA}$ in Pb+Pb and Xe+Xe collisions.
  The dashed curves show the results for $\kappa=2.55$, corresponding
  to the mixed
  scenario with mQGP formation in $pp$ collisions only
  at the LHC energies.
          Data points are from PHENIX \cite{PHENIX_r}.
}
\end{figure}
%

The results for $R_{pp}$, obtained for the optimal value
$\kappa=2.55$ for $c=0.8$, are shown in Fig.~3 for
the RHIC energy $\sqrt{s}=0.2$, and the LHC energies $\sqrt{s}=2.76$,
$5.02$, and $7$ TeV.
To illustrate sensitivity of the results to the value of $\tau_0$,
we show in Fig.~3 the results for $\tau_0=0.8$ fm. As one can see,
the difference between the results for $\tau_0=0.5$ and $0.8$ fm
is rather small, especially for the LHC energies.
The curves for $\tau_0=0.8$ fm in Fig.~ 3 are obtained with the optimal
$\kappa$ for $R_{AA}$ for $\tau_0=0.5$ fm. The results for $R_{pp}$
obtained for the value of $\kappa$ adjusted from fit of $R_{AA}$
with $\tau_0=0.8$ fm are very close to that shown in Fig.~3
for $\tau_0=0.5$ fm.
The curves shown in Fig.~3 correspond to our fit of the heavy
  ion data with the geometrical parameter $k=2$. Calculations
  of $R_{pp}$ with $\kappa\approx 2.98$ obtained from the heavy ion fit
  with $k=3$ give a small ($\lsim 5$\%) decrease of $1-R_{pp}$.
  Thus, our results for $R_{pp}$ are quite stable against variation of the
  parameter $k$ used for calculations of $R_{AA}$ in heavy ion collisions.
Note that the specific form (\ref{eq:120}) of the fireball
radius $R_f$ from the IP-Glasma model, used in our calculations,
is practically unimportant for $R_{pp}$, because, at a given UE
multiplicity density, it varies very slowly
with $R_f$. We observed that at $p_T\sim 10-20$ GeV the variation of $R_{pp}$
with the fireball radius
for the interval $R_f/1.3-R_f1.3$ (with $R_f$ given by (\ref{eq:130}))
is $\lsim 2$\%, and at $p_T\gsim 50$ GeV it becomes $\lsim 0.3$\%.
Note that the $R_{pp}$ obtained in the present work is closer to 1 than
the band for $R_{pp}$ given in our previous analysis \cite{Z_pp13}.
There calculations were performed for a temperature independent
$\alpha_s^{fr}$, invariable in the whole QGP fireball.
This entails some arbitrariness in the predictions for $R_{pp}$ as
different choices of $\alpha_s^{fr}$ for large and small size QGPs
are possible.

In Figs.~4--6 we compare our results for $R_{AA}$ for charged hadrons
with the LHC data for 2.76 and 5.02 TeV Pb+Pb and 5.44 TeV
Xe+Xe collisions used in our $\chi^2$ fitting.
We show the curves for the optimal values of $\kappa$
for the scenarios with and without mQGP formation. In Figs.~4--6
  we show the results for the
  geometrical parameter $k=2$.
For a given $\kappa$ obtained with $k=2$, $R_{AA}$ at $k=3$
decreases by
$\sim 5-10$\% at $p_T\sim 10-20$ GeV and by $\sim 1-3$\%
at $p_T\sim 50-100$ GeV as compared to $R_{AA}$ for $k=2$.
The difference between
$R_{AA}$ for $k=2$ and $k=3$ for the optimal values of $\kappa$ 
is very  small (typically $\lsim 1-2$\%)  at all
$p_T$.
Therefore we do not show the curves for $k=3$.
From Figs.~4--6 one can see that for optimal values of $\kappa$
the difference between theoretical $R_{AA}$ for our two
scenarios is very small.
From Figs.~4--6 one can see that the theoretical curves
for the scenarios with and without mQGP formation in $pp$ collisions
agree quite well with experimental data. 
Our results for $R_{AA}$ obtained for $\tau_0=0.8$ fm
(not shown) turn out to be very close to that for $\tau_0=0.5$ fm.
Even if $R_{AA}$ for $\tau_0=0.8$ fm is calculated with
the optimal value of $\kappa$ obtained for $\tau_0=0.5$ fm,
it grows just by $\sim 1-2$\%. This says that jet quenching
in heavy ion collisions is not very sensitive
to the initial stage of the fireball evolution.
The results of our calculations for $c=0$ in the
parametrization (\ref{eq:80}) are very close to that for $c=0.8$
(shown in Figs. 4--6), therefore we do not plot them as well.

In Fig.~7 we compare the results for $R_{AA}$ of $\pi^0$-mesons
in $0.2$ TeV Au+Au collisions with the data from PHENIX \cite{PHENIX_r}.
The theoretical curves are obtained with the optimal values
of $\kappa$ adjusted by fitting the LHC data.
Note that a joint fit of the RHIC and the LHC data
practically does not change the optimal values of $\kappa$ obtained from
the LHC data, because the number of the
data points for the LHC data set is much bigger than for the PHENIX data.
From Fig.~7 one can see that the theoretical curves somewhat overshoot
the data at $p_T\sim 10$ GeV. However, the discrepancy between the
results for RHIC and the LHC for the $T$-dependent
$\alpha_s$ is much smaller than in the case of calculations
with a unique, temperature independent, $\alpha_s$ \cite{RAA20}.
Since conditions for formation of a mQGP  in $pp$ collisions
are less favorable than for the LHC energies,
it is interesting to compare the RHIC data on $R_{AA}$
also for an intermediate scenario when mQGP
formation in $pp$ collisions occurs only at the LHC energies.
In this case the theoretical $R_{AA}$ for $0.2$ TeV Au+Au collisions
should be calculated without $1/R_{pp}$ factor for the optimal
$\kappa$ fixed in the scenario with mQGP formation
from the LHC data on $R_{AA}$.
In Fig.~7 we show the results for this case,
using $\kappa=2.55$ obtained in the scenario with mQGP formation
at the LHC energies. As one can see, it gives somewhat better
agreement with the PHENIX data \cite{PHENIX_r}. This fact supports
the intermediate scenario without a fully developed hydrodynamic
evolution of the QCD matter in $0.2$ TeV $pp$ collisions.
Of course, this interpretation of the discrepancy between $\alpha_s$ for the
RHIC and the LHC energies is quite speculative, because
our calculations include several simplifications, and
the errors for the PHENIX data on $R_{AA}$ at $p_T\gsim 10$ GeV
are rather large.

\begin{figure}[!h]
\begin{center}
\includegraphics[height=8.5cm]{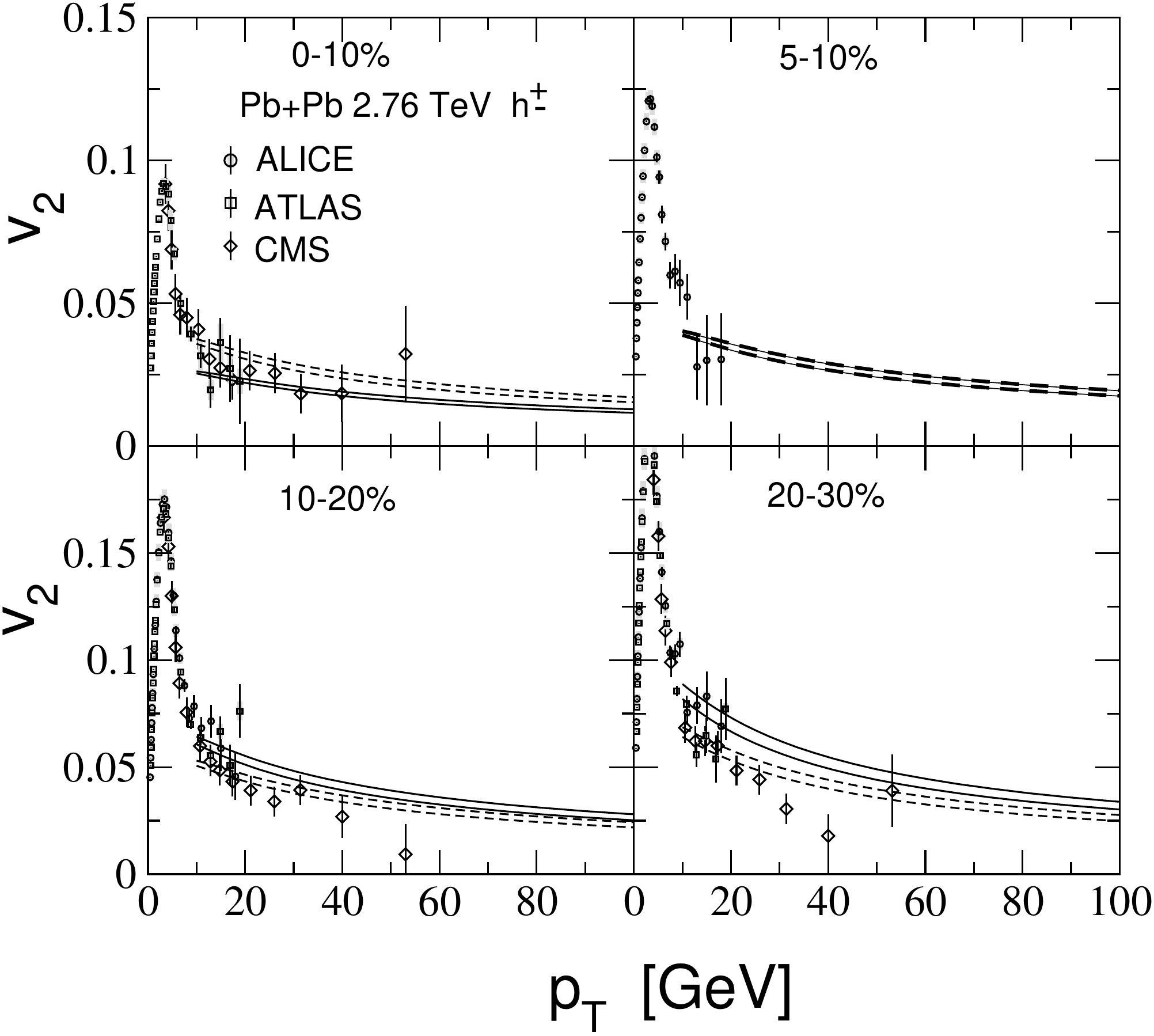}  
\end{center}
\caption[.]
        {$v_2$ of charged hadrons in $2.76$ TeV  Pb+Pb collisions
for the initial fireball eccentricity $\epsilon_2$ calculated 
in the optical (solid) and Monte-Carlo (dashed) Glauber model
for scenarios without and with (bottom to top) mQGP formation in
$pp$ collisions. The parameters are the same as for Fig. 4.
Data points are from ALICE \cite{ALICE_v276}, ATLAS \cite{ATLAS_v276},
  and CMS \cite{CMS_v276}.}
\end{figure}
\begin{figure}[!h]
\begin{center}
\includegraphics[height=8.5cm]{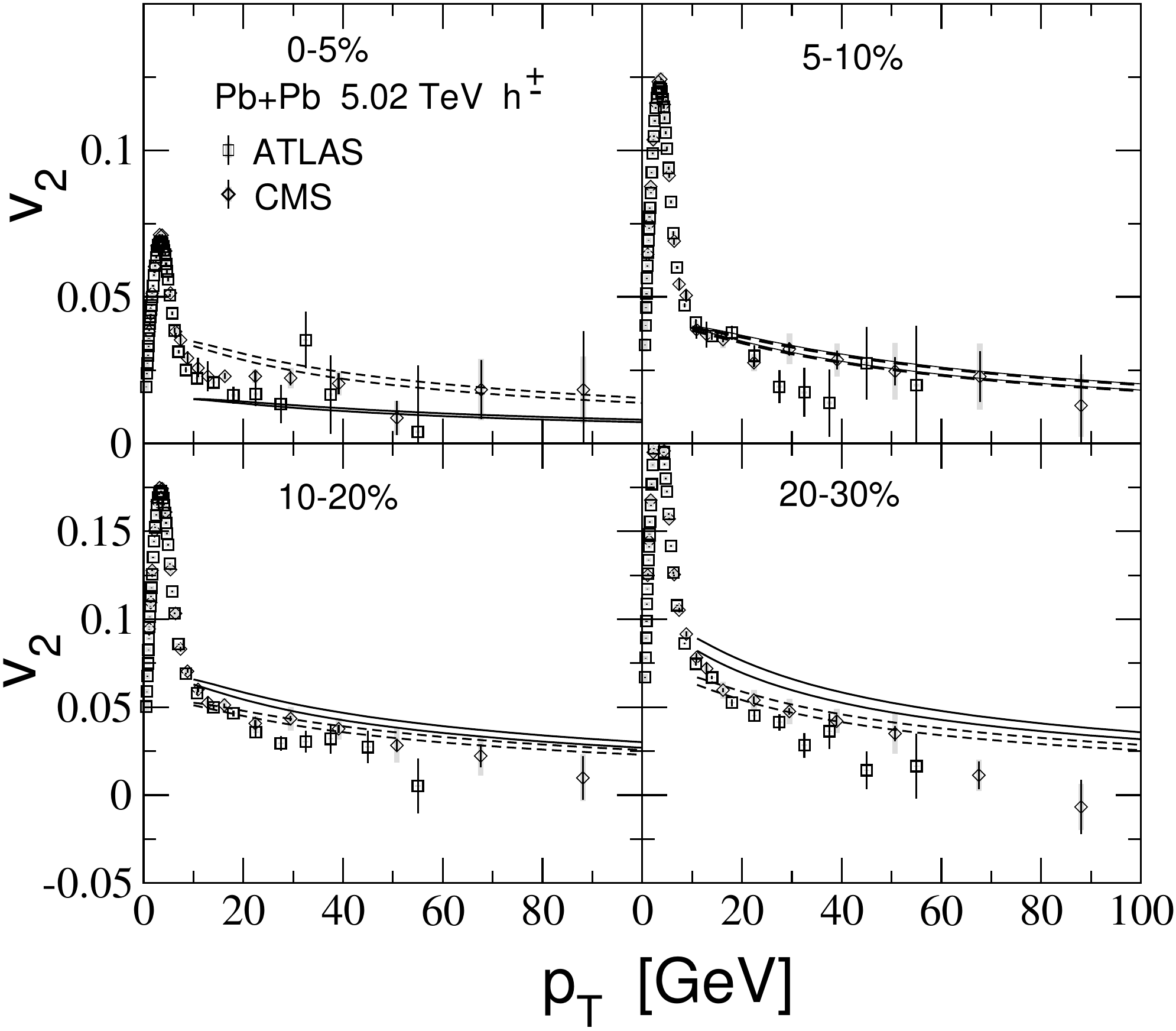}  
\end{center}
\caption[.]
        {Same as in Fig.~8 for $\sqrt{s}=5.02$ TeV.
The parameters are the same as for Fig. 5.          
Data points are from ATLAS \cite{ATLAS_v502} and CMS \cite{CMS_v502}.
}
\end{figure}

In Figs.~8 and 9 we compare the theoretical predictions
for the flow coefficient $v_2$ with available
experimental data for $2.76$ \cite{ALICE_v276,ATLAS_v276,CMS_v276} 
and $5.02$ \cite{ATLAS_v502,CMS_v502} TeV Pb+Pb collisions.
We show the theoretical results for the fireball azimuthal
asymmetry $\epsilon_2$ obtained in the optical Glauber model
and for the Monte-Carlo one (for the Gaussian entropy sources
with width $\sigma=0.7$ fm). The most significant difference
in predictions of the optical and the Monte-Carlo
versions is for the most central (0-5\%) collisions.
Although $v_2$ was not included into our $\chi^2$ analysis, the
results for $v_2$ are in qualitative agreement with
what is seen in the LHC data. Of course, one should bear in mind,
that results for the azimuthal dependence are less robust.
For instance, for $v_2$, which is sensitive to a delicate balance between
jet modification in the directions along the short and long fireball
semi-diameters, the errors due to the neglect of the transverse
QGP motion and due to the use of a flat initial entropy distribution
can potentially be larger than for the azimuthally averaged $R_{AA}$.
Note that for $v_2$ the difference in the theoretical
predictions for scenarios with and without mQGP is more pronounced
than for $R_{AA}$.
It is due to some enhancement of the strength of jet quenching
for the scenario with mQGP formation (due to smaller value
of $\kappa$) as compared to the one without mQGP.
Contrary to the case of $R_{AA}$, for $v_2$ this difference is now not
compensated by the presence of the $1/R_{pp}$ factor, because
this factor is immaterial for calculation of the azimuthal
asymmetry.

\begin{figure}[!h]
\begin{center}
\includegraphics[height=8.5cm]{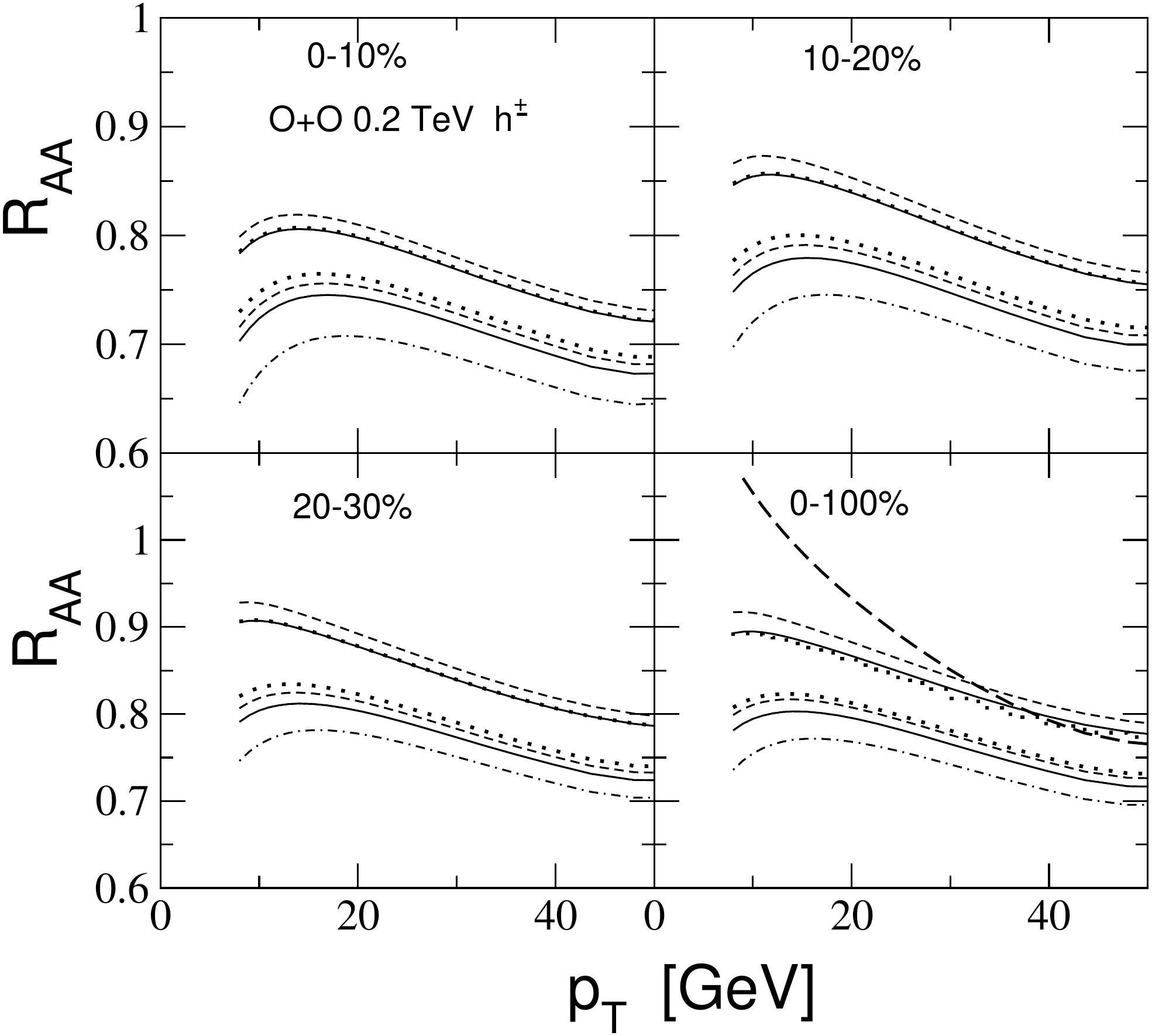}  
\end{center}
\caption[.]
        {
$R_{AA}$ of charged hadrons for $0.2$ TeV O+O collisions
  from our calculations
  for scenarios (top to bottom) with ($\kappa=2.55$) and without
($\kappa=3.47$) mQGP formation in $pp$
collisions. 
The solid and dashed curves are for $\tau_0=0.5$, and the dotted ones
are for $0.8$ fm.
  The solid and dotted curves are obtained
  for the geometrical parameter $k=2$, and the dashed curves
are  for $k=3$ (see text for explanation).
The dash-dotted lines show the results (for
$k=2$, $\tau_0=0.5$ fm, $\kappa=2.55$)
  for the intermediate scenario when mQGP formation in $pp$ collisions
  occurs at the LHC energies, but it
  is absent at RHIC. 
  The long dashed line shows
  $R_{AA}^{pdf}$.         }
\end{figure}
\begin{figure}[!h]
\begin{center}
\includegraphics[height=8.5cm]{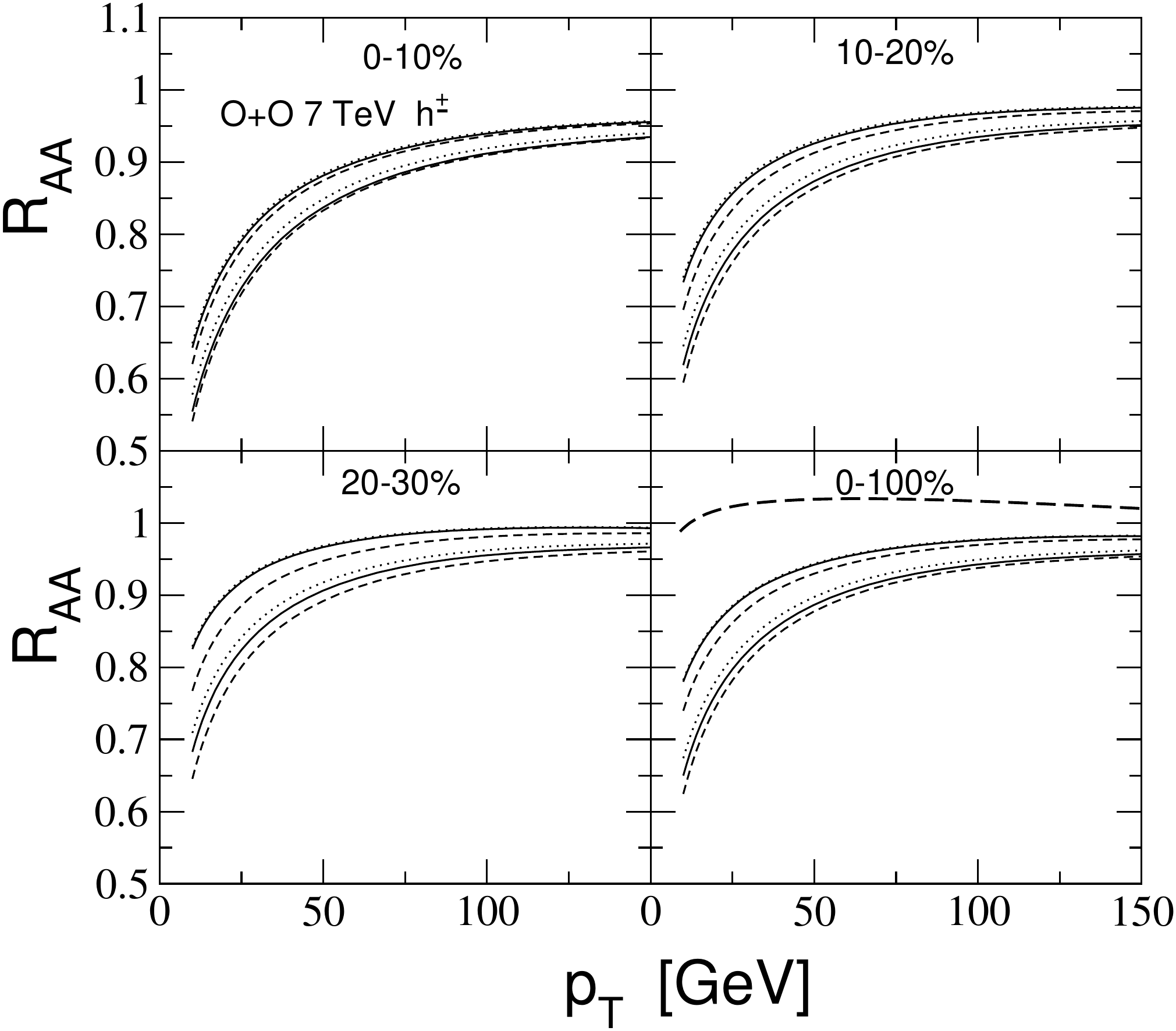}  
\end{center}
\caption[.]
        {
          Same as in Fig.~10 (but now without dash-dotted lines)
          for $7$ TeV O+O collisions.
}
\end{figure}
\begin{figure}[!h]
\begin{center}
\includegraphics[height=8.5cm]{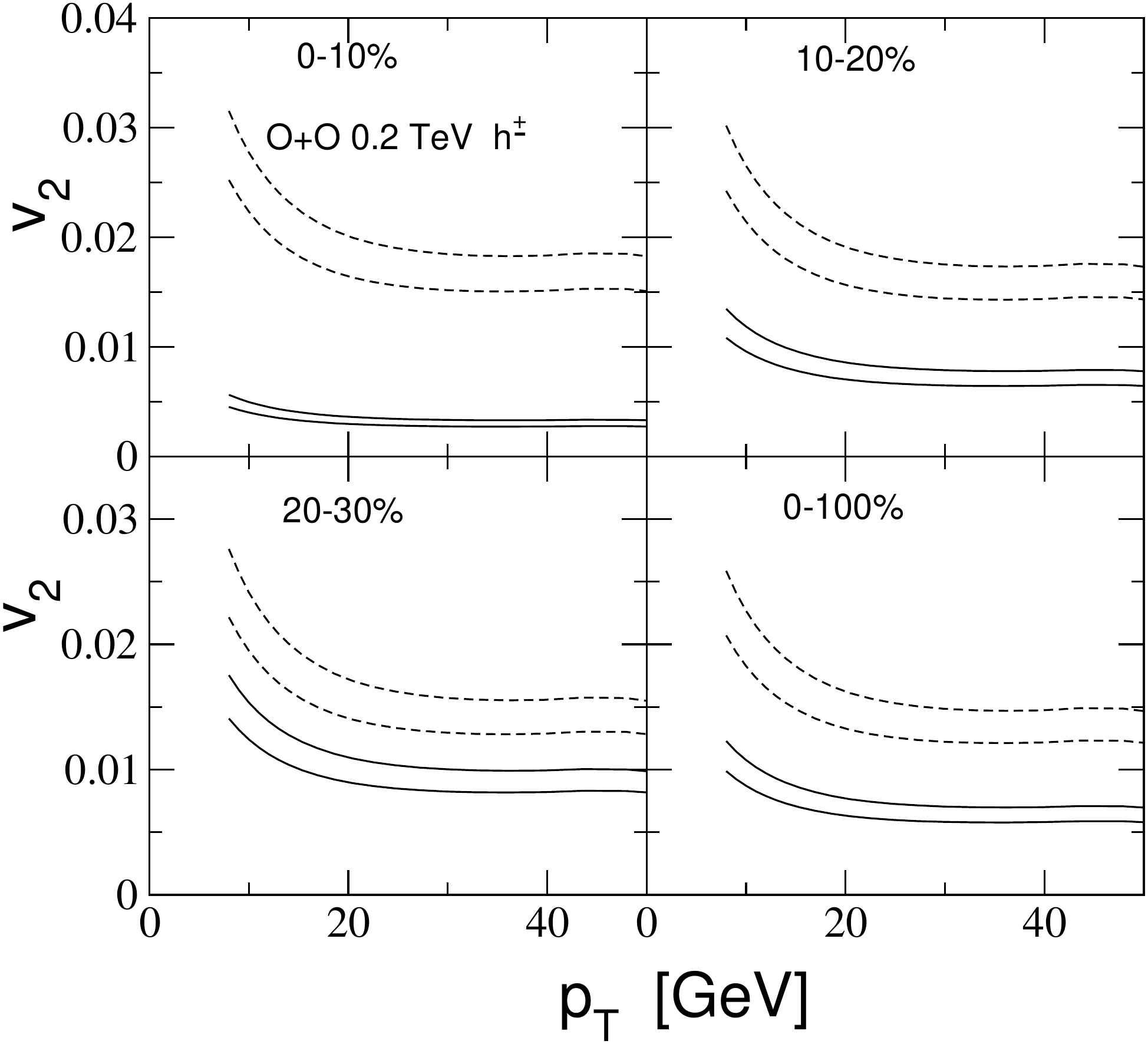}  
\end{center}
\caption[.]
        {$v_2$ of charged hadrons in $0.2$ TeV O+O collisions
for $\tau_0=0.5$ for the initial fireball eccentricity $\epsilon_2$ calculated 
in the optical (solid) and Monte-Carlo (dashed) Glauber model
for scenarios without and with (bottom to top) mQGP formation
in $pp$ collisions. 
}
\end{figure}
\begin{figure}[!h]
\begin{center}
\includegraphics[height=8.5cm]{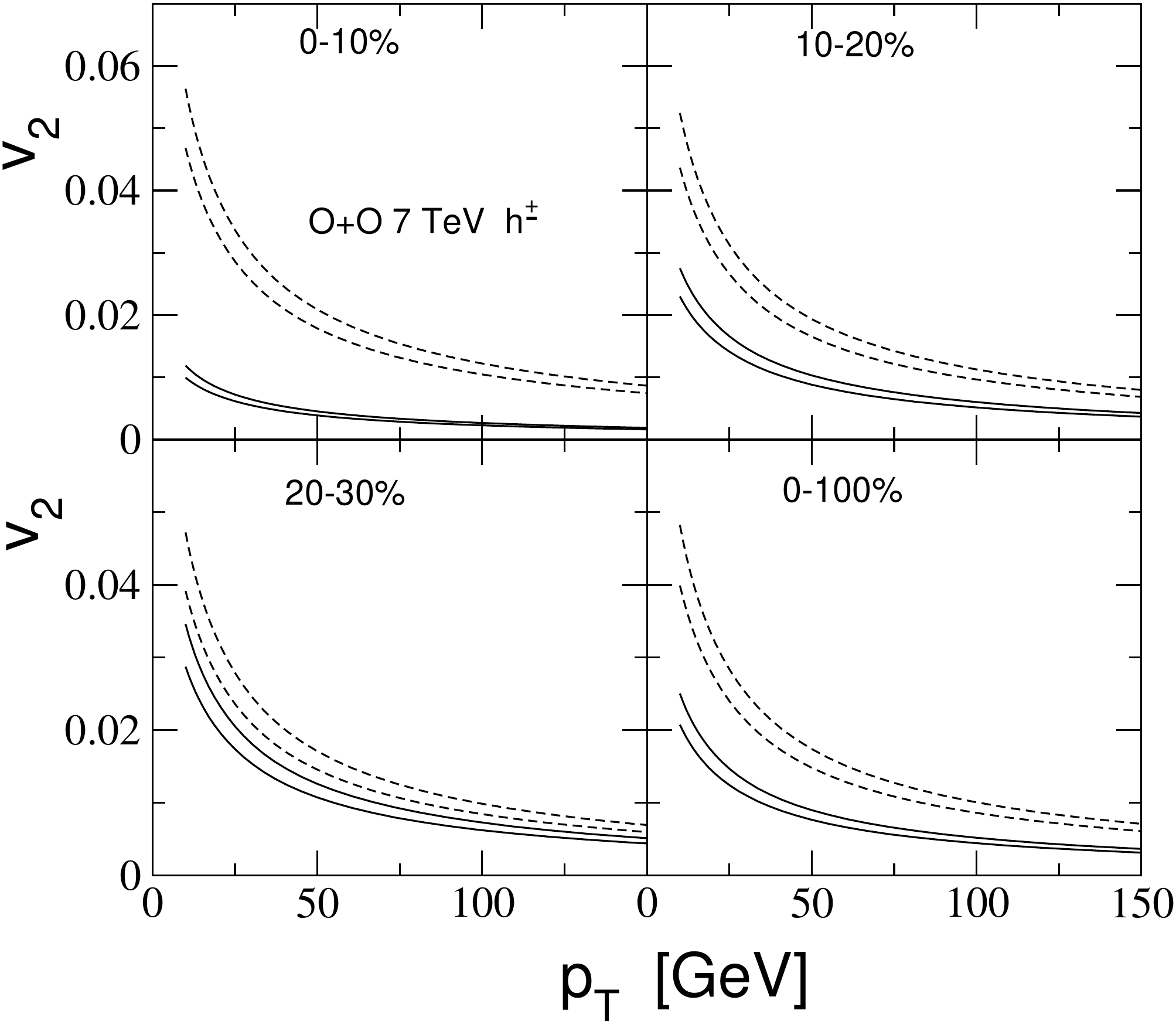}  
\end{center}
\caption[.]
        {Same as in Fig.~12 for $\sqrt{s}=7$ TeV.
}
\end{figure}
Thus, we have established that the scenarios with and without mQGP
can describe equally reasonably jet quenching in heavy ion collisions.
Now, using the optimal values of $\kappa$, adjusted to fit
$R_{AA}$ for heavy ion collisions, we can calculate
$R_{AA}$ for O+O collisions.
In Figs.~10 and 11 we show the results for $R_{AA}$ in $0.2$ and $7$ TeV
O+O collisions for scenarios with and without mQGP formation
in $pp$ collisions. We present results for three narrow centrality bins
and for the minimum bias $R_{AA}$ (i.e., for $0-100$\% centrality interval).
The latter quantity is particularly convenient for comparison
with experiment,
because for light ion collisions, due to large multiplicity fluctuations,
the impact parameter becomes less correlated with centrality, which is defined
experimentally via charged multiplicity.
For $\sqrt{s}=0.2$ TeV, as in Fig.~7
for $0.2$ TeV Au+Au collisions, we also
show predictions for the intermediate scenario with
mQGP formation in $pp$ collisions only at the LHC energies.
As one can see, this scenario leads to the smallest $R_{AA}$
for $0.2$ TeV O+O collisions.
To illustrate sensitivity of the results to $\tau_0$, in Figs.~10 and 11
we present the results for $\tau_0=0.5$ and $0.8$ fm.
As one can see, the difference between predictions
for these two values of $\tau_0$ is rather small, especially
for $\sqrt{s}=7$ TeV.
In Figs.~10 and 11, in addition to the real $R_{AA}$, we also plot
the modification factor for hadron spectra only due to
nuclear modification of the parton distribution
functions  for $AA$ collisions (which we denote $R_{AA}^{pdf}$).
For light ion collisions, when $R_{AA}\sim 1$, it is reasonable
to characterize magnitude of jet quenching by $\Delta R_{AA}=R_{AA}^{pdf}-R_{AA}$.
From Figs.~10 and 11 we observe that,
 both for $\sqrt{s}=0.2$ and $7$ TeV, there is a substantial
difference in $\Delta R_{AA}$ for scenarios with and without mQGP
formation in $pp$ collisions. Thus, although the discrepancy
between $R_{AA}$ for these two scenarios are very small for heavy
ion collisions, they give very different $\Delta R_{AA}$ for
light ion collisions.
To illustrate dependence of the theoretical predictions
on the choice of the nuclear geometry parameter $k$, which we use to define
the overlap region of the colliding nuclei, in Figs.~10 and 11, in addition
to the version with $k=2$, we also plot the results for $k=3$.
As one can see, there is no a significant difference
between calculations for $k=2$ and $3$.
From Figs.~10 and 11 we can conclude that
possible variation of $R_{AA}$ in O+O collisions  under change of the
parameters $k$ and $\tau_0$ turns out to be smaller than 
the difference between predictions
for our two scenarios without and with mQGP formation in $pp$ collisions.

From Fig.~11 one can see that for $7$ TeV O+O collisions
the minimum bias $\Delta R_{AA}$
for scenarios without and with mQGP formation
may differ by a factor of $\sim 1.5- 2$ at $p_T\sim 50-150$ GeV.
Nevertheless, the $p_T$-dependence of $R_{AA}$ for
both models are similar. Also, one should bear in mind  that
$R_{AA}^{pdf}$ has its own theoretical uncertainties. All this can make
it difficult to discriminate
between the scenarios without and with mQGP formation
from comparison with the future LHC data.
From Fig.~10 we observe that for $0.2$ TeV O+O collisions
the difference between scenarios with and without mQGP
formation is somewhat more pronounced than at $\sqrt{s}=7$ TeV.
It is interesting that in this case the minimum bias $R_{AA}$ for
the mQGP scenario becomes larger $R_{AA}^{pdf}$ at $p_T\gsim 35$ GeV.
An observation of such a crossover would
support mQGP formation in $pp$ collisions.

In Figs.~12 and 13 we plot the results for the
flow coefficient $v_2$ for the initial fireball azimuthal
asymmetry $\epsilon_2$ obtained within the optical Glauber
wounded nucleon model and within its Monte-Carlo counterpart.
Similarly to  $v_2$ for Pb+Pb collisions, shown in Figs.~8 and 9,
we obtain somewhat larger $v_2$ for scenario with 
mQGP formation. As already said above, our results for $v_2$ in O+O
collisions are of illustrative character, because for O+O collisions
the effect of fluctuations of the initial entropy density
is very important. For this reason our purpose
is just to demonstrate the relation between the initial fireball
ellipticity $\epsilon_2$ and $v_2$. From Fig.~13 we observe
that for $7$ TeV O+O collisions $\epsilon_2/v_2\sim 20-25$ at
$p_T\sim 50$ GeV.

\subsection{Jet quenching in $pA$ collisions}
We have not touched yet jet quenching in $pA$ collisions.
Similarly to formula (\ref{eq:30}), the theoretical observable nuclear
modification factor for $pA$ collisions is given by
$R_{pA}=R_{pA}^{st}/R_{pp}$.
Unfortunately, a conclusive comparison with experiment is presently
impossible, because the available data on $R_{pA}$
are somewhat contradictory. ALICE measurement \cite{ALICE_RPA5}
gives the minimum bias $R_{pA}$ of charged hadrons
in 5.02 TeV $p$+Pb collisions  which is $\sim 0.92-1$
at $p_T\gsim 10$. More recent ALICE measurement \cite{ALICE_RPA8} of $R_{pA}$
for $\pi^0$ mesons in $8.16$ TeV $p$+Pb collisions also
gives $R_{pA}$ which is close to unity at $p_T\gsim 10$ GeV.
In contrast, the CMS  collaboration \cite{CMS_RPA5}  
obtained for charged hadrons in 5.02 TeV $p$+Pb collisions
$R_{pA}\sim 1.1-1.19$ at $p_T\gsim 10$ GeV.
The LHC data from ALICE and CMS seem to exclude the scenario when
mQGP formation and jet quenching are absent in $pp$ collisions
(i.e., $R_{pp}\approx 1$), but are present in $pA$ collisions (i.e., $R_{pA}<1$).
If one assumes that the data from ALICE \cite{ALICE_RPA5,ALICE_RPA8}
are correct, the question arises whether
the scenario with mQGP formation, both in $pp$ and $pA$ collisions,
can be consistent with $R_{pA}$ measured by ALICE.
It is possible if the theoretical nuclear modification factor $R_{pA}^{st}$
is just a little smaller than $R_{pp}$.
Calculations of $R_{pA}^{st}$ require detailed information on the
QGP fireball in $pA$ collisions. We have performed preliminary
calculations of $R_{pA}$ for 5.02 and 8.16 TeV $p$+Pb collisions
using the fireball parameters obtained by modeling the UE fireball
within the Monte-Carlo Glauber wounded nucleon model.
The important difference
between the minimum bias events and the jet events in the wounded nucleon
Glauber picture is that in the jet events always there is one 
wounded $pN$ pair participating in the hard parton production,
which produces the UE multiplicity
density $dN_{ch}^{ue}(pp)/d\eta$ enhanced by the $K_{ue}\sim 2.5$ factor,
as compared to the minimum bias multiplicity density $dN_{ch}^{mb}(pp)/d\eta$.
This leads to a two component structure of the fireball, with a high
density core and the low density peripheral part (related to
spectator wounded nucleons, that do not participate in the hard process).
Our Monte-Carlo simulation shows that for $p$+Pb collisions
the peripheral fireball part 
can give $\sim 20-30$\% of the total UE
multiplicity density $dN_{ch}^{ue}(pA)/d\eta$.
Since the entropy density in the peripheral part
is rather small (the ideal gas temperature $\sim 150-170$ MeV),
one can expect that it should be close to the free-streaming regime,
and its effect on jet quenching should be small.
The core region of the UE fireball in 5.02(8.16) TeV $p$+Pb collisions,
which can have collective behavior,
generates the charged multiplicity density $\sim 23\pm 3(27\pm 3)$.
Calculations with such charged multiplicity densities
give $R_{pA}=R_{pA}^{st}/R_{pp}$ which agrees quite well with
the ALICE data \cite{ALICE_RPA5,ALICE_RPA8}.
Of course, this analysis is of preliminary character.
It is highly desirable to investigate the jet quenching effect
of the peripheral part of the fireball (that can be close to the free-streaming
regime) and to perform calculations for a realistic
fireball density profile accounting for the transverse flow (although
we expect that the effect of the density variation and the flow corrections
should be small (see section 2)).
We leave this, more complicated, calculations for
future work.
%


\section{Conclusions}
In the present work, we have studied jet quenching
for  heavy and light ion collisions
within the LCPI  approach \cite{LCPI1} to induced
gluon emission. Calculations of the induced gluon spectrum are
performed with accurate treatment
of the Coulomb effects, using the method suggested in \cite{RAA04}.
We account for multiple gluon emission in the approximation
of independent gluon radiation \cite{RAA_BDMS} in the form suggested
in \cite{RAA08} (with
some modifications made in \cite{RAA20}).
We perform calculations for the temperature dependent
running QCD coupling.
We use parametrization of $\alpha_s(Q,T)$ which has a plateau
around $Q=Q_{fr}\sim \kappa T$
(it is motivated by the lattice calculation of the effective
QCD coupling in the QGP \cite{Bazavov_al1} and calculations
within the functional renormalization group \cite{RG1}).

We have investigated scenarios with and without
mQGP formation in $pp$ collisions, and an intermediate scenario
with mQGP formation only at the LHC energies.
The parameter $\kappa$ has been fitted to
the LHC data on the nuclear modification factor $R_{AA}$
in $2.76$ and $5.02$ TeV Pb+Pb, and $5.44$ TeV Xe+Xe collisions.
We find that predictions of the models with and without mQGP formation
in $pp$ collisions are very similar, and both of them lead to quite good
description of the LHC heavy ion data on $R_{AA}$. The theoretical
predictions for the flow coefficient $v_2$ are also in reasonable
agreement with the LHC data. 
The optimal values of $\kappa$, fitted to the LHC data
on $R_{AA}$, lead to reasonable description of
the PHENIX data \cite{PHENIX_r} on $R_{AA}$ in $0.2$ TeV Au+Au collisions
in scenarios with and without mQGP formation in $pp$ collisions.
Agreement with the PHENIX data becomes somewhat better
for the intermediate scenario with mQGP formation
in $pp$ collisions only at the LHC energies.
Overall, our analysis shows that the experimental data on jet quenching in
heavy ion collisions can equally well be described
in scenarios without and with mQGP formation in $pp$ collisions.

Using the optimal values of the parameter $\kappa$, obtained by fitting $R_{AA}$
in heavy collisions, we perform calculations of $R_{AA}$ and $v_2$
for 0.2 and 7 TeV O+O collisions which 
may be studied in future experiments at RHIC \cite{CERN_STAR}
and at the LHC \cite{CERN_yel,CERN_LHC,CERN_OOsum}.
We find that predictions of scenarios with and without
mQGP formation for $R_{AA}-R_{AA}^{pdf}$
begin to differ substantially for O+O collisions.
However, due to theoretical uncertainties for $R_{AA}^{pdf}$,
and the fact that the $p_T$-dependence of $R_{AA}$ for
both scenarios are similar, it may be difficult to discriminate
between the scenarios without and with mQGP formation
from comparison with the future LHC data.
We find that for $0.2$ TeV O+O collisions
the difference between scenarios with and without mQGP
formation in $pp$ collisions is somewhat more pronounced.
Our calculations show that in this case
the minimum bias $R_{AA}$ for the mQGP scenario
becomes larger $R_{AA}^{pdf}$ at $p_T\gsim 35$ GeV.

For the medium modification factor $R_{pp}$ (that is not directly
observable quantity) we obtain at $p_T=10$ GeV
$R_{pp}\sim 0.84-0.88$ at $\sqrt{s}=0.2$ TeV,
and $R_{pp}\sim 0.77-0.82$ at the LHC energies.

\begin{acknowledgments}
  I am grateful to S.~Tripathy
  for useful communication on some aspects of the recent
ALICE results on jet quenching in small collision systems   
\cite{ALICE_Ipp}. 	
  This work was performed under the Russian Science Foundation grant 20-12-00200
at Steklov Mathematical Institute.
\end{acknowledgments}

\end{document}